\documentclass{aastex}
\usepackage{emulateapj5}
\usepackage{graphicx}
\shorttitle{MHD Turbulence}
\shortauthors{Cho, Lazarian \& Vishniac}
\begin{document}
\title{Ordinary and Viscosity-Damped MHD Turbulence}
\author{Jungyeon Cho and A. Lazarian}
\affil{Dept. of Astronomy, University of Wisconsin,
   Madison, WI53706; cho, lazarian@astro.wisc.edu}
\and
\author{Ethan T. Vishniac}
\affil{Physics and Astronomy Dept., Johns Hopkins University, 3400 N. Charles St.,
       Baltimore, MD 21218; ethan@pha.jhu.edu }

\newcommand{\be}{\begin{equation}}
\newcommand{\ee}{\end{equation}}
\newcommand{\bea}{\begin{eqnarray}}
\newcommand{\eea}{\end{eqnarray}}

\begin{abstract}
We compare the properties 
of ordinary strong magnetohydrodynamic (MHD) turbulence 
in a strongly magnetized
medium with the recently discovered viscosity-damped regime. 
We focus on  energy spectra, anisotropy,  and intermittency.
Our most surprising conclusion is that in ordinary strong MHD
turbulence the velocity and magnetic fields show
different high-order structure function scalings. Moreover this
scaling depends on whether the intermittency is viewed in a global
or local system of reference. This reconciles
seemingly contradictory earlier results.
On the other hand, the intermittency scaling for viscosity-damped 
turbulence is very different, and difficult to understand in terms
of the usual phenomenological models for intermittency in turbulence.
Our remaining results are in reasonable agreement with expectations.
First, we find that our high resolution 
simulations for ordinary MHD turbulence show that the energy spectra are
{\it compatible} with a Kolmogorov spectrum, while
viscosity-damped turbulence shows a shallow $k^{-1}$ spectrum for 
the magnetic fluctuations.  Second, a new numerical technique confirms 
that ordinary MHD turbulence exhibits Goldreich-Sridhar type anisotropy,
while viscosity-damped MHD turbulence shows extremely anisotropic eddy 
structures. Finally, we show that
many properties of incompressible turbulence for both the
ordinary and viscosity-damped regimes carry over to the 
case of compressible turbulence.

\end{abstract}


%%%%%%%%%%%%%%%%%%%%%%%%%%%%%%%%%%%%%%%%%%%%%%%%%%%%%%%%%%%%%%%%%%%%%

\section{Introduction}

The interstellar medium (ISM) shows  
density and velocity statistics that indicate
the existence of strong turbulence 
(Armstrong, Rickett, \& Spangler 1995; Lazarian \& Pogosyan 2000;
Stanimirovic \& Lazarian 2001; see also the review by
Cho, Lazarian, \& Vishniac 2003 and references therein).
Understanding interstellar turbulence is essential for many astrophysical
processes including 
star formation, and  cosmic ray transport (see reviews by 
Vazquez-Semadeni et al. 2000; Lazarian, Cho, Yan 2002 and references
therein). Moreover, the importance of magnetohydrodynamic (MHD) 
turbulence is not limited to interstellar processes. For instance,
properties of MHD turbulence may be essential for describing gamma
ray bursts (see Lazarian et al. 2003) and magnetic reconnection
(Lazarian \& Vishniac 1999). 

Kolmogorov theory (1941) is the simplest model for incompressible
hydrodynamic turbulence.
The main prediction of the theory is the velocity-separation relation
$v_l \sim l^{1/3}$, or energy spectrum $E(k)\sim k^{-5/3}$, where
$l$ is the separation between two points (or, we regard it as eddy size) and
$k\sim 1/l$ is the wave number. In general, $kE(k) \sim v_l^2$.
Astrophysical turbulence differs from its laboratory counterpart in
that it is usually magnetized, compressible, and the gas can be partially 
ionized.  However, the single most important difference is that typical
large eddy scales in MHD turbulence are usually many orders of magnitude
larger than any relevant microphysical scale.  This allows us to 
sidestep many of the issues that make laboratory plasma physics so 
complicated.

Including magnetic field effects has a dramatic effect on strong
turbulence.  Our current understanding of MHD turbulence
owes a great deal to fundamental results obtained 
by Shebalin, Matthaeus, \& Montgomery  (1983), Higdon (1984),
Matthaeus et al. (1998) and other researchers (see
a more complete list in Cho, Lazarian \& Vishniac 2003). Recently 
Goldreich \& Sridhar (1995; hereafter GS95) 
proposed a model for incompressible MHD turbulence,
which was later supported by numerical simulations (Cho \& Vishniac 2000; 
Maron \& Goldreich 2001).
The GS95 model predicts a Kolmogorov energy spectrum, $E(k)\sim k^{-5/3}$,
for both velocity and magnetic fields.
The major difference between MHD and hydrodynamic turbulence is 
the anisotropy of the eddy structures -
eddies are statistically isotropic in hydrodynamic turbulence 
while eddies show scale-dependent anisotropy 
(i.e. smaller eddies are more anisotropic) in the MHD case
(see Cho, Lazarian, \& Vishniac 2003).

How useful is studying incompressible turbulence?
Does it have anything to do with turbulence in realistic compressible
flows?
Compressible MHD turbulence has been studied intensively by
different researchers (for a review see Cho \& Lazarian 2003c and
references therein). A discussion of the extent to which the incompressible
GS95 scaling is applicable to realistic compressible fluids may
be found in the original GS95 paper.  More recently
Lithwick \& Goldreich (2001) extended GS95 model for high-$\beta$ 
($\beta\equiv P_{gas}/P_{magnetic}$) plasmas, i.e. 
for the case when gas pressure is larger than magnetic pressure.
They also made a conjecture about
the scaling of slow modes in low-$\beta$ plasmas. 
Cho \& Lazarian (2002) studied turbulence in low-$\beta$ plasmas and
obtained numerical scaling relations for both compressible and Alfvenic
modes. Finally, both the high- and low-$\beta$ plasma cases are covered
in Cho \& Lazarian (2003a,b). They showed that the Alfvenic and slow modes
follow the GS95 scaling not only for Mach numbers much less than 
unity, but also for much larger Mach numbers, as long as the motions
were still sub-Alfvenic. Fast modes were
shown to be decoupled from the rest of the modes\footnote{Prior to this
research it was erroneously believed that the linear modes are strongly
coupled in the case of MHD turbulence and this was claimed to be the
reason for a rapid decay of turbulent motions. A recent study 
by Vestuto, Ostriker \& Stone (2003) confirmed our findings on a
marginal coupling of compressible and incompressible MHD motions.} 
and isotropic.  The work on compressible turbulence shows that 
incompressible simulations are meaningful
in terms of representing most features of realistic flows.
Consequently in this paper we concentrate on incompressible turbulence.

In this paper, we focus on the effects of viscosity.
In strong hydrodynamic turbulence energy is injected at a scale $L$,
and cascades down to smaller scales without significant viscous losses
until it reaches the viscous damping scale $l_{dv}$.
The Kolmogorov energy spectrum applies to the inertial range, i.e.
all scales between $L$ and $l_{dv}$.
This simple picture becomes more complicated when we deal with MHD turbulence 
because there are two dissipation scales - the velocity damping scale
$l_{dv}$ and the magnetic diffusion scale $l_{dm}$, where magnetic
structures are dissipated.
%-----------
%In fully ionized collisionless plasmas (e.g. the hot ionized medium), 
%both scales are comparable and small.
In fully ionized collisionless plasmas (e.g. the hottest phases of
the ISM), $l_{dv}$  is less than an order of magnitude
larger than $l_{dm}$, but both scales are very small.
%-----------
However, in partially ionized plasmas (e.g. the warm or cold neutral phase
of the ISM), the two dissipation scales are very different and 
$l_{dv}\gg l_{dm}$.
In the Cold Neutral Medium (see Draine \& Lazarian 1999 for a list of
the idealized phases) neutral particle transport leads to viscous
damping on a scale which is a fraction of a parsec. 
In contrast, in these same phases $l_{dm}\sim 100km$.

This has a dramatic effect on the energy cascade model in 
a partially ionized medium.
When the energy reaches the viscous damping scale $l_{dv}$,
kinetic energy will dissipate there, but the magnetic
energy will not.
In the presence of dynamically important magnetic field,
Cho, Lazarian, \& Vishniac (2002b; hereafter CLV02b) reported 
a completely new regime of turbulence below the scale at which
viscosity damps kinetic motions of fluids\footnote{
   Further research showed that there is a smooth connection 
   between this regime and small scale turbulent dynamo in high 
   Prandtl number fluids (see Schekochihin et al. 2002). 
}. 
They showed that
magnetic fluctuations extend below the viscous damping scale
and form a shallow spectrum $E_b(k)\sim k^{-1}$. 
%-----------
The spectrum is similar to that of the viscous-convective range of
a passive scalar in hydrodynamic turbulence (see, for example,
Lesieur 1990).
%-----------
In addition they showed
that turbulence in the new regime is very anisotropic and intermittent.
Here, we compare ordinary and viscosity-damped MHD turbulence.
We mainly focus on the energy spectra, eddy anisotropy,  
scale-dependent intermittency, and intermittency from
high order velocity statistics.

In what follows, we briefly
consider a theoretical model for viscosity-damped
turbulence (\S 2). A detailed theoretical study is given in 
Lazarian, Vishniac \& Cho (2003; hereafter LVC03). 
We discuss our numerical method
in \S 3. We present our results on anisotropies and intermittency 
in \S 4, while results on higher-order statistics are given in
\S 5. A discussion and the summary are presented in \S 6 and \S 7 
respectively.

%%%%%%%%%%%%%%%%%%%%%%%%%%%%%%%%%%%%%%%%%%%%%%%%%%%%%%%%%%%%%%%%%%%%%

\section{A Theoretical Model for Viscosity-Damped MHD Turbulence}

Following the usual treatment of ordinary strong MHD turbulence, we define
the wavenumbers $k_{\|}$ and $k_{\perp}$ as the components of the
wavevector measured along the {\it local}
mean magnetic field and perpendicular to it, respectively.  In this   
Here the local mean magnetic field is the direction of the locally averaged
magnetic field, which depends not only on the location but also
the volume over which the average is taken.
See Cho \& Vishniac (2000) and Cho, Lazarian, \& Vishniac 
(2002a; hereafter CLV02a) for details.

Lazarian, Vishniac, \& Cho (LVC03) 
proposed a theoretical model for
viscosity-damped MHD turbulence.
We summarize the  model as follows.

Since there is no significant velocity fluctuation below $l_{dv}$,
the time scale for the energy cascade below $l_{dv}$
is fixed at the viscous damping scale.  Consequently
the energy cascade time scale $t_{cas}$ 
is scale-independent below $l_{dv}$ and
the requirement for a scale independent energy rate rate $b_l^2/t_{cas}$ yields
\begin{equation}
  b_l \sim \mbox{constant, or~~~} E_b(k)\sim k^{-1},
\end{equation}
where $k E_b(k)\sim b_l^2$.

In LVC03, we assume that the curvature of the magnetic field lines
changes slowly, if at all, in the cascade:
\begin{equation}
  k_{\|} \sim \mbox{constant}.
\end{equation}
  This is consistent
with a picture in which the cascade is driven by repeated shearing
at the same large scale.  It is also consistent with the numerical
work described in CLV02b, which yielded a constant $k_{\|}$
throughout the viscously damped nonlinear cascade.  A corollary
is that the wavevector component in the direction of the perturbed
field is also approximately constant, so that the increase in $k$ is
entirely in the third direction.

The kinetic spectrum depends on the scaling of intermittency.
In LVC03, we define  a filling factor $\phi_l$, which is the
fraction of the volume containing strong magnetic field perturbations
with a scale $l \sim k^{-1}$.  We denote the velocity and perturbed
magnetic field inside these subvolumes with a ``$\hat{\ } $'' so
that
\begin{equation}
v_l^2=\phi_l \hat v_l^2,
\end{equation}
and
\begin{equation}
b_l^2=\phi_l \hat b_l^2.
\end{equation}
We can balance viscous and magnetic tension forces
to find
\begin{equation}
{\nu\over l^2} \hat v_l \sim \max[\hat b_lk_c,B_0k_{\|,c}] \hat b_l
\sim k_c\hat b_l^2,
\label{s3}
\end{equation}
where $k_c \sim 1/l_{dv}$
%-----------------------------------
and $k_{\|,c}$ is the parallel component of the wave vector corresponding
to the perpendicular component $k_c$.
We used the Goldreich-Sridhar scaling ($B_0k_{\|,c}\sim b_lk_c$)
   and $\hat b_l \geq b_l$ to evaluate the two terms in the square braces.
%-----------------------------------
Motions on scales smaller than $l_{dv}$ will be 
continuously sheared at a rate $\tau_s^{-1}$.
These structures will reach a dynamic equilibrium if they generate a
comparable shear, that is
\begin{equation}
{\hat v_l\over l}\sim \tau_s^{-1} \sim \mbox{constant}.
\label{s1}
\end{equation}
Combining this with equation (\ref{s3}), we get
\begin{equation}
 \phi_l\sim {k_c l}
\end{equation}
and
\begin{equation}
 E_v(k) \sim k^{-4}.
\end{equation}
Note that equation (\ref{s3}) implies that
kinetic spectrum would be $ E_v(k) \sim k^{-5}$ if $\phi_l$=constant.

%%%%%%%%%%%%%%%%%%%%%%%%%%%%%%%%%%%%%%%%%%%%%%%%%%%%%%%%%%%%%%%%%%%%%

\section{Method}
\subsection{Numerical Method}
We have calculated the time evolution of incompressible magnetic turbulence
subject to a random driving force per unit mass.
We have adopted a pseudo-spectral code to solve the
incompressible MHD equations in a periodic box of size $2\pi$:
\begin{equation}
\frac{\partial {\bf v} }{\partial t} = (\nabla \times {\bf v}) \times {\bf v}
      -(\nabla \times {\bf B})
        \times {\bf B} + \nu \nabla^{2} {\bf v} + {\bf f} + \nabla P' ,
        \label{veq}
\end{equation}
\begin{equation}
\frac{\partial {\bf B}}{\partial t}=
     \nabla \times ({\bf v} \times{\bf B}) + \eta \nabla^{2} {\bf B} ,
     \label{beq}
\end{equation}
\be
      \nabla \cdot {\bf v} =\nabla \cdot {\bf B}= 0,
\ee
where $\bf{f}$ is a random driving force,
$P'\equiv P/\rho + {\bf v}\cdot {\bf v}/2$, ${\bf v}$ is the velocity,
and ${\bf B}$ is magnetic field divided by $(4\pi \rho)^{1/2}$.
In this representation, ${\bf v}$ can be viewed as the velocity 
measured in units of the r.m.s. velocity
of the system and ${\bf B}$ as the Alfven speed in the same units.
The time $t$ is in units of the large eddy turnover time ($\sim L/V$) and
the length in units of $L$, the scale of the energy injection.
In this system of units, the viscosity $\nu$ and magnetic diffusivity $\eta$
are the inverse of the kinetic and magnetic Reynolds numbers respectively.
The magnetic field consists of the uniform background field and a
fluctuating field: ${\bf B}= {\bf B}_0 + {\bf b}$.
We use 21 forcing components with $2\leq k \leq \sqrt{12}$, where
wavenumber $k$ is in units of $L^{-1}$.
Each forcing component has correlation time of one.
The peak of energy injection occurs at $k\approx 2.5 $.
The amplitudes of the forcing components are tuned to ensure $v \approx 1$
We use exactly the same forcing terms
for all simulations.
The Alfv\'en velocity of
the background field, $B_0$, is set to 1.
In pseudo spectral methods, the temporal evolution of
equations (\ref{veq}) and (\ref{beq}) are followed in Fourier space.
To obtain the Fourier components of nonlinear terms, we first calculate
them in real space, and transform back into Fourier space.
The average helicity in these simulations is not zero. However,
previous tests have shown that our results are insensitive to the value of the 
kinetic helicity.
In incompressible fluid dynamics $P'$ is not an independent variable.
We use an appropriate projection operator to calculate
$\nabla P'$ term in
{}Fourier space and to enforce the divergence-free condition
($\nabla \cdot {\bf v} =\nabla \cdot {\bf B}= 0$).
We use up to $256^3$ collocation points.
We use an integration factor technique for kinetic and magnetic dissipation terms
and a leap-frog method for nonlinear terms.
We eliminate the $2\Delta t$ oscillation of the leap-frog method by using
an appropriate average.
At $t=0$, the magnetic field has only its uniform component
and the velocity field is restricted to the range
$2\leq k \leq 4$ in wavevector space.

For the ordinary turbulence, we mostly use 8-th order hyper-viscosity and
hyper-diffusion, so that 
the viscosity and diffusion terms in the equations (\ref{veq}) and (\ref{beq})
become
\be
 -\nu_8 (\nabla^2)^8{\bf v} \mbox{~~~and~~~} -\eta_8 (\nabla^2)^8{\bf B},
\ee
respectively.
Here, $\nu_8 ~(=\eta_8)$ is adjusted in such a way that
the dissipation cutoff occurs right before $k\sim N/3$, 
where $N$ is the number of grids in each spatial direction.
This way, we can avoid the aliasing error
of pseudo-spectral method.

For viscosity-damped turbulence, 
we mostly use
a physical viscosity ($\nu=0.015$) and a third order 
hyper-diffusion for magnetic field,
so that the dissipations in the equation (\ref{beq})
is replaced with
\be
 -\eta_3 (\nabla^2)^3 {\bf B}.
\ee
However, for the run 256PP-B$_0$1, we use 
a physical diffusion for magnetic field.

We list parameters used for the simulations in Table 1.
The run 384PH3-B$_0$1 is exactly the same as the run with the same name in 
CLV02b.
We use the notation 384XY-$B_0$Z,
where 384 refers to the number of grids in each spatial
direction; X, Y = P, H3, H8 refers to physical or hyper-diffusion 
(and its power);
Z=0, 1 refers to the strength of the external magnetic field.

\subsection{Parameter Space}
We require that,
at the energy injection scale, $B_0 k_{\|,L} \approx V_L k_{\perp, L}$.
Therefore, we take $B_0 \sim 1$ for most of the runs (see Table 1).

For ordinary MHD turbulence, the dissipation cutoff ($k_c \sim 1/l_d$) 
occurs at $k\sim N/3$.
For viscosity-damped MHD turbulence,
we require that $k_c \sim k_L$ ($\sim 1/L$) to maximize the dynamical range.
For 384PH3-B$_0$1, we use $\nu \sim 0.015$ (or, Reynolds number $R \sim 100$)
to guarantee turbulence.
In fact, the energy spectrum shows that the viscous cutoff occurs at $k\sim 7$ when
we take $\nu \sim 0.015$.
When we take $\nu \sim 0.06$ (e.g. run 256PH8-B$_0$0.5),
the viscous cutoff occurs right at the energy injection scale $k_L$.
In this case, we do not expect a turbulent velocity field.

%%%%%%%%%%%%%%%%%%%%%%%%%%%%%%%%%%%%%%%%%%%%%%%%%%%%%%%%%%%%%%%%%%%%%
\section{Spectra, anisotropy, and scale-dependent intermittency}

\subsection{Spectra and $k_{\|}$}
\begin{figure*}
  \includegraphics[width=0.49\textwidth]{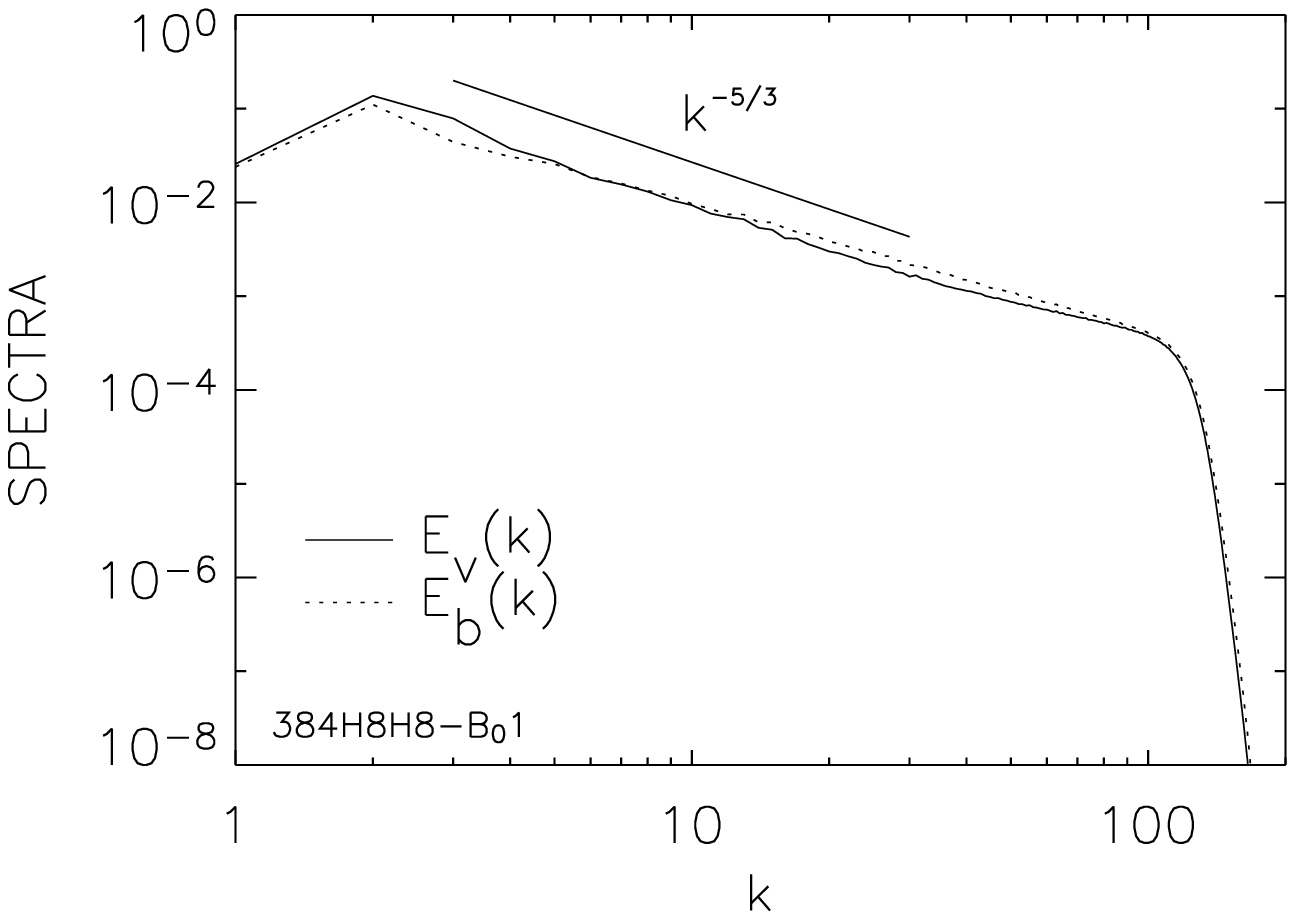}
\hfill
  \includegraphics[width=0.49\textwidth]{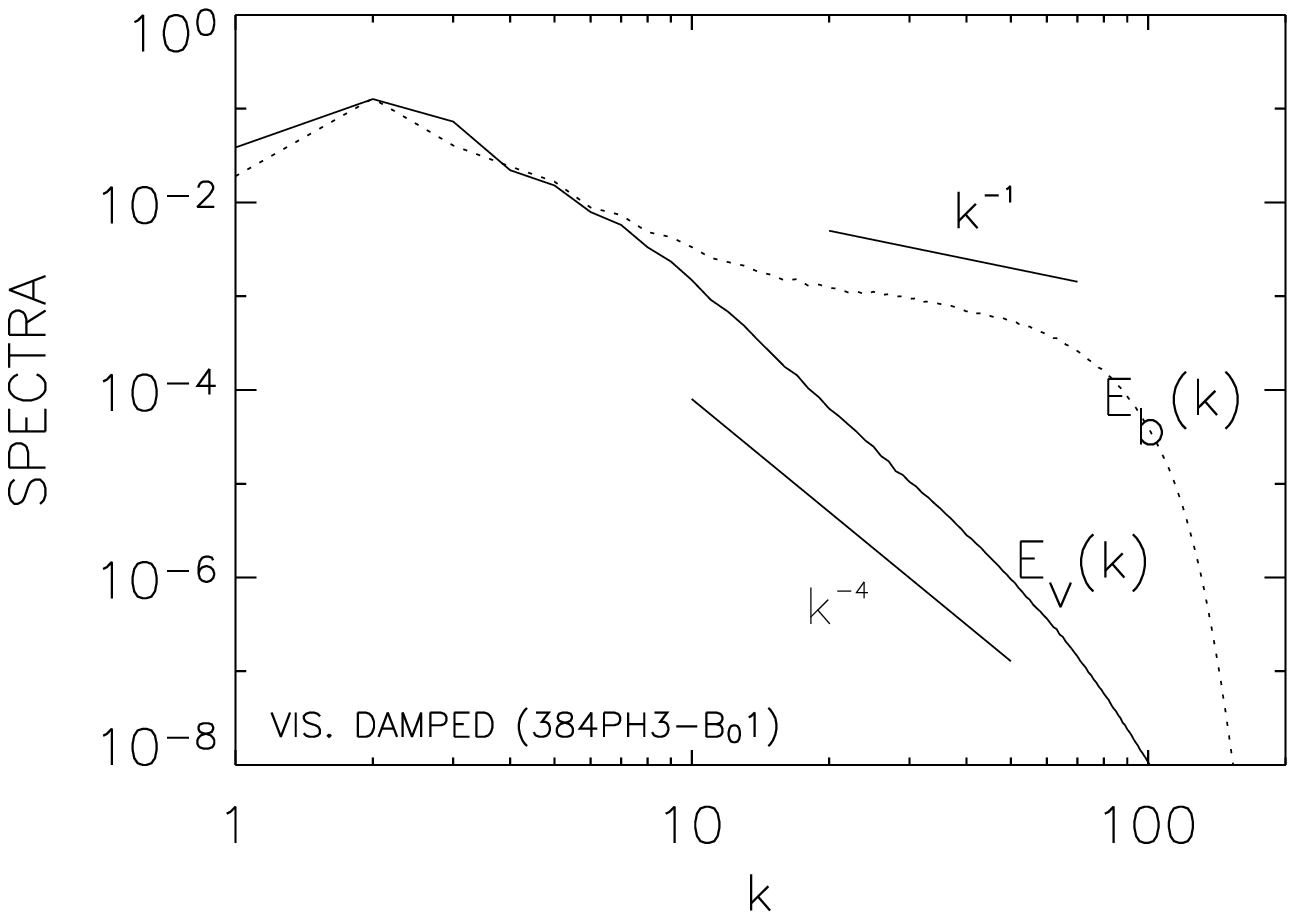}
  \caption{ Spectra.
   (a: {\it left}) Ordinary MHD turbulence (384H8H8-B$_0$1). 
       The kinetic spectrum ($E_v(k)$) is compatible with the Kolmogorov
       spectrum. However, the magnetic spectrum ($E_b(k)$) 
       is slightly shallower than
       Kolmogorov.
   (b: {\it right}) Viscosity-damped MHD turbulence (384PH3-B$_0$1).
       The kinetic spectrum declines quickly after 
       the viscous cutoff at $k\sim 7$.
       The magnetic spectrum follows a $k^{-1}$ power-law after the cutoff,
       in agreement with theoretical expectations.
       The slope of $E_v(k)$ is somewhat steeper than the $k^{-4}$
       dependence that follows from theoretical considerations. {}Figure (b) is
similar to that in CLV02b.
}
\label{fig_sp}
\end{figure*}

Figure~\ref{fig_sp} shows the energy spectra of ordinary and 
viscosity-damped turbulence.
In both cases, energy is injected at $k\sim 2.5$.
In ordinary turbulence, the injected energy cascades down
to the single scale at $k\sim 100$.
In the viscosity-damped case, the kinetic energy and magnetic energy cascade
together to  the viscous damping scale at $k\sim 7$.
Beyond $k\sim 7$, kinetic spectrum drops sharply due to viscous damping.
In contrast the magnetic spectrum flattens out to 
a $k^{-1}$ spectrum for $10<k<80$.

%%%%%%%%%%%%%%%%%%%%%%%%%%%%%%%%% spectra
%%%%%%%%%%%%%%%%%%%%%%%%%%%%%%%%% spectra

\begin{figure*}
  \includegraphics[width=0.49\textwidth]{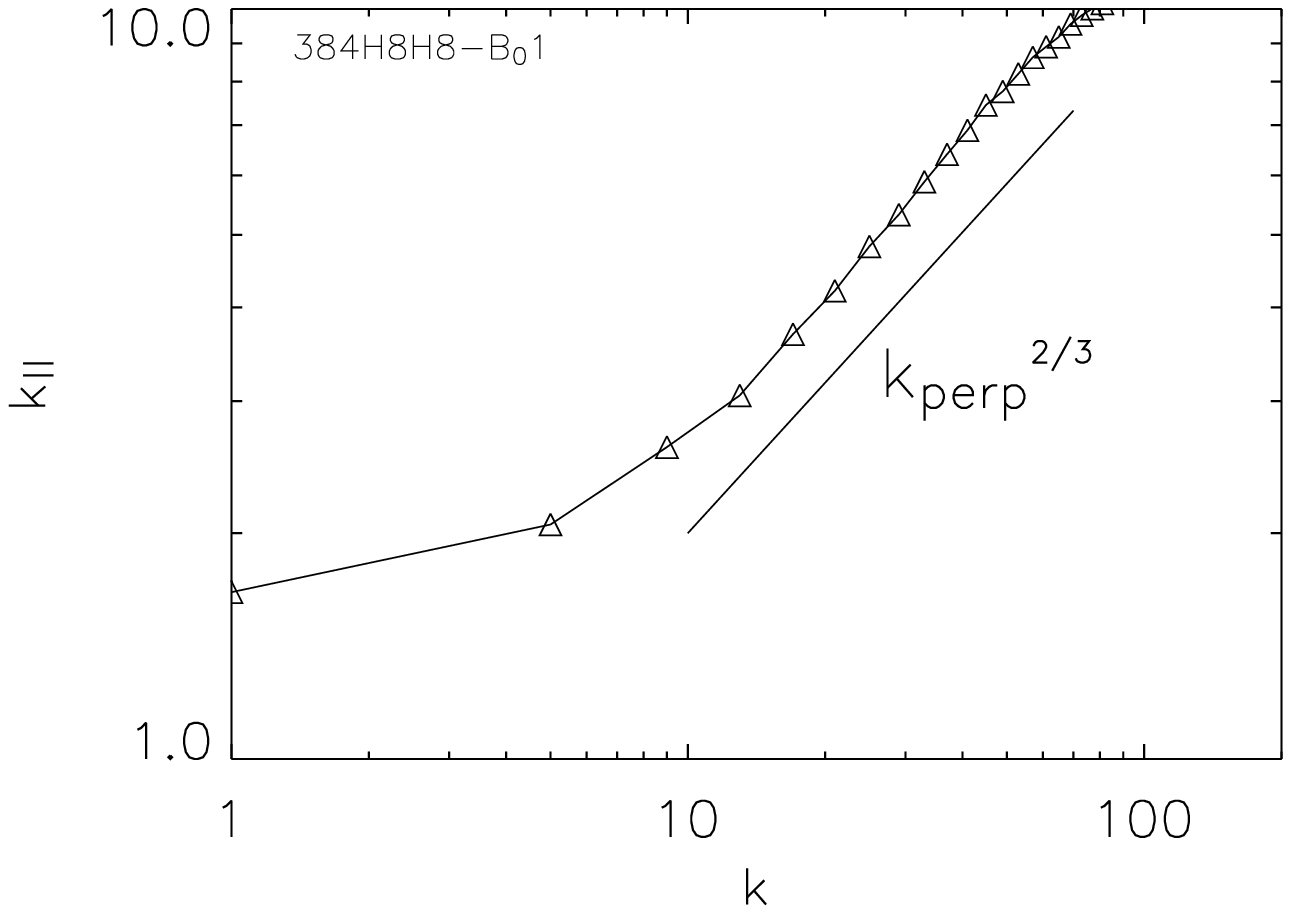}
\hfill
  \includegraphics[width=0.49\textwidth]{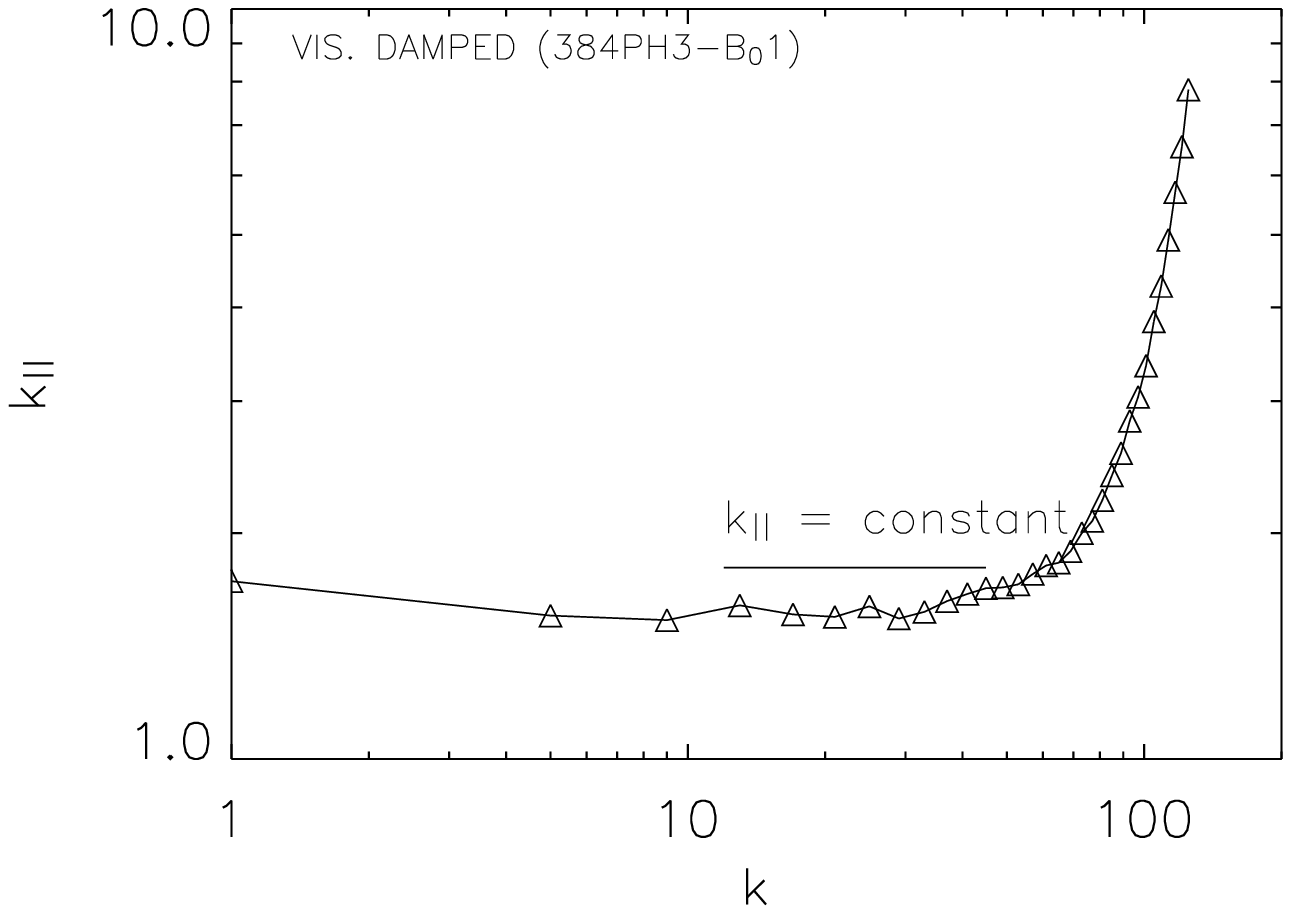}
  \caption{ Scaling of parallel wavenumber $k_{\|}$, which is a measure of
anisotropy.
   (a: {\it left}) Ordinary MHD.
       This result confirms the scale-dependent anisotropy prediction of
       Goldreich \& Sridhar (GS95): $k_{\|}\sim k_{\perp}^{2/3}$.
   (b: {\it right}) Viscosity-damped MHD.
       The parallel wavenumber $k_{\|}$ remains almost constant
       in the viscosity-damped regime ($k>10$). {}Figure (b) is
       similar to one in CLV02b. 
}
\label{fig_kpar}
\end{figure*}

Figure~\ref{fig_kpar} shows the parallel wave number $k_{\|}$
as a function of the total wavenumber.
A method for calculating $k_{\|}$ is described in CLV02b.
The term ${\bf B}\cdot \nabla {\bf B}$ describes magnetic tension
and is approximately equal to $B_0 k_{\|} b_l$ for an isolated eddy
in a uniform mean field ${\bf B}_0$. 
Here $k_{\|} \propto 1/l_{\|}$ and $l_{\|}$
is the characteristic length scale parallel to ${\bf B}_0$, which
is known to be larger than the perpendicular length scale.
%%%
Cho \& Vishniac (2000) and
CLV02a argued that, in actual turbulence, 
eddies are aligned with the local mean field ${\bf B}_L$.
%%%
We can obtain the {\it local frame representation} of
$k_{\|}$, by considering an eddy lying 
in the {\it local} mean field ${\bf B}_L$: 
${\bf B}_L\cdot \nabla {\bf b}_l \approx B_L k_{\|} b_l.$\footnote{
There can be many ways to define the local mean field ${\bf B}_L$.
In CLV02b and this paper, 
we obtain ${\bf B}_L$ for an eddy of (perpendicular) 
size $l\propto 1/k$
  by eliminating
modes whose perpendicular wavenumber is greater than $k/2$ and
  ${\bf b}_l$  by eliminating
modes  whose perpendicular wavenumber is less than $k/2$.
}
{}The Fourier transform of this relation yields
%%%\be
$
 |\widehat{ {\bf B}_L\cdot \nabla {\bf b}_l }|_{\bf k}
%%%------------------
    \approx B_L k_{\|} |\hat{\bf b}|_{\bf k},
%%%------------------
$
%%%\ee
where hatted variables are Fourier-transformed quantities.
{}From this, we have
\be
k_{\|} \approx \left( 
     \frac{\sum_{k\leq |{\bf k}^{\prime}| <k+1} 
   |\widehat{ {\bf B}_L\cdot \nabla {\bf b}_l   }|_{{\bf k}^{\prime}}^2  }
  { B_L^2 \sum_{k\leq |{\bf k}^{\prime}| <k+1} 
%%%------------------
                     |\hat{\bf b}|^2_{{\bf k}^{\prime}} }
%%%------------------
               \right)^{1/2}.    \label{kparLl}
\ee

Figure \ref{fig_kpar}(a) shows that, for {\it ordinary} turbulence, 
$k_{\|}$ measured by this method
gives results consistent with the GS95 relations 
between $k_{\|}$ and  $k_{\perp}$:  $k_{\|} \sim k_{\perp}^{2/3}$.
This new method is complementary to the previous
method utilizing
structure functions (Cho \& Vishniac 2000) for the study of
anisotropy.
Figure \ref{fig_kpar}(b) shows that,
for {\it viscosity-damped} turbulence, $k_{\|}\approx$
constant.

%%%%%%%%%%%%%%%%%%%%%%%%%%%%%%%%%%%%% kpar
%%%%%%%%%%%%%%%%%%%%%%%%%%%%%%%%%%%%% kpar

\subsection{Scale-dependent intermittency}
The theoretical model in LVC03 predicts that
intermittency is scale-dependent for viscosity-damped MHD turbulence.
For ordinary MHD turbulence, no theory has addressed this issue.

\begin{figure*}
\begin{tabbing}
\includegraphics[width=0.49\textwidth]{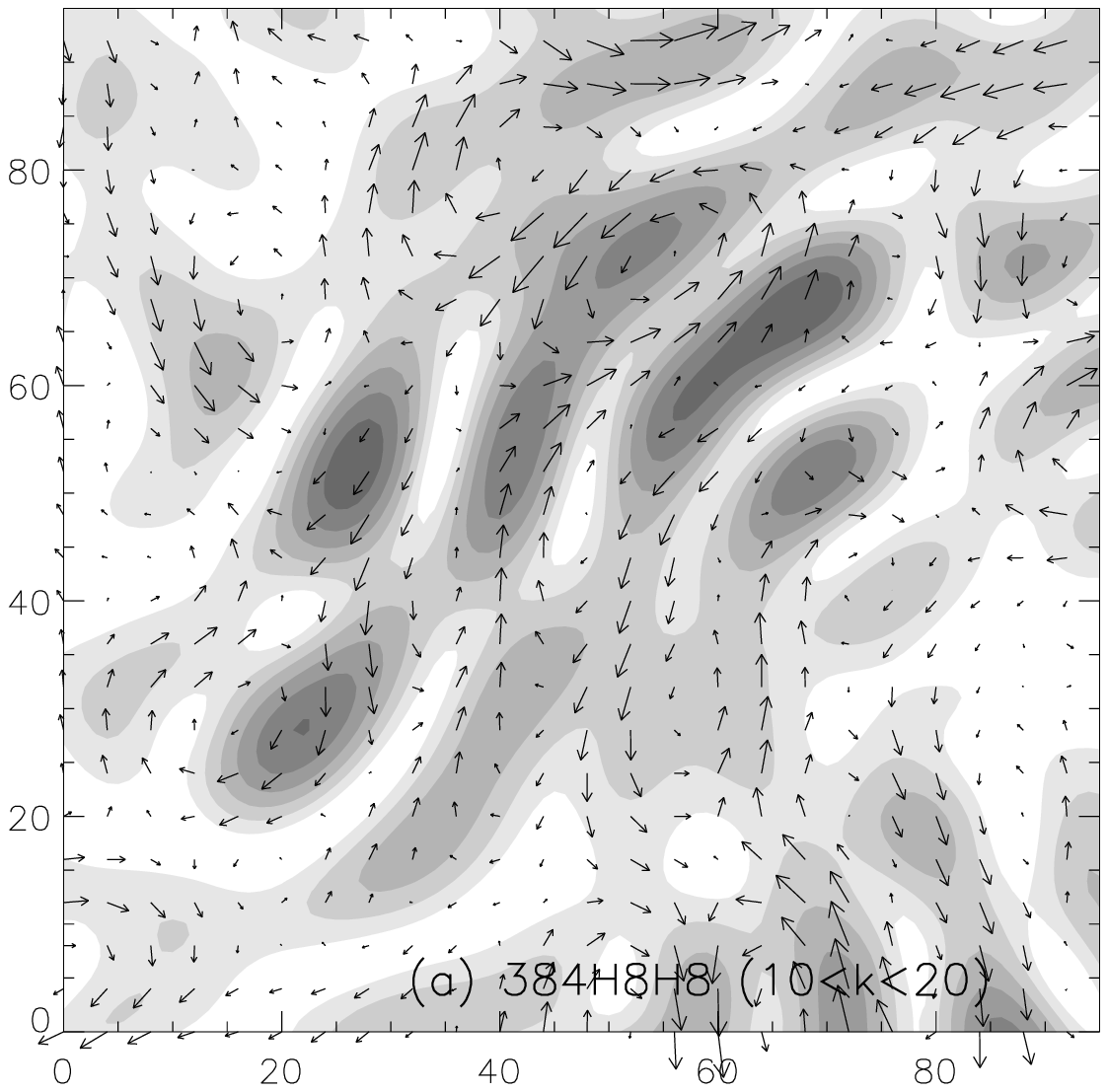}
\=
\includegraphics[width=0.49\textwidth]{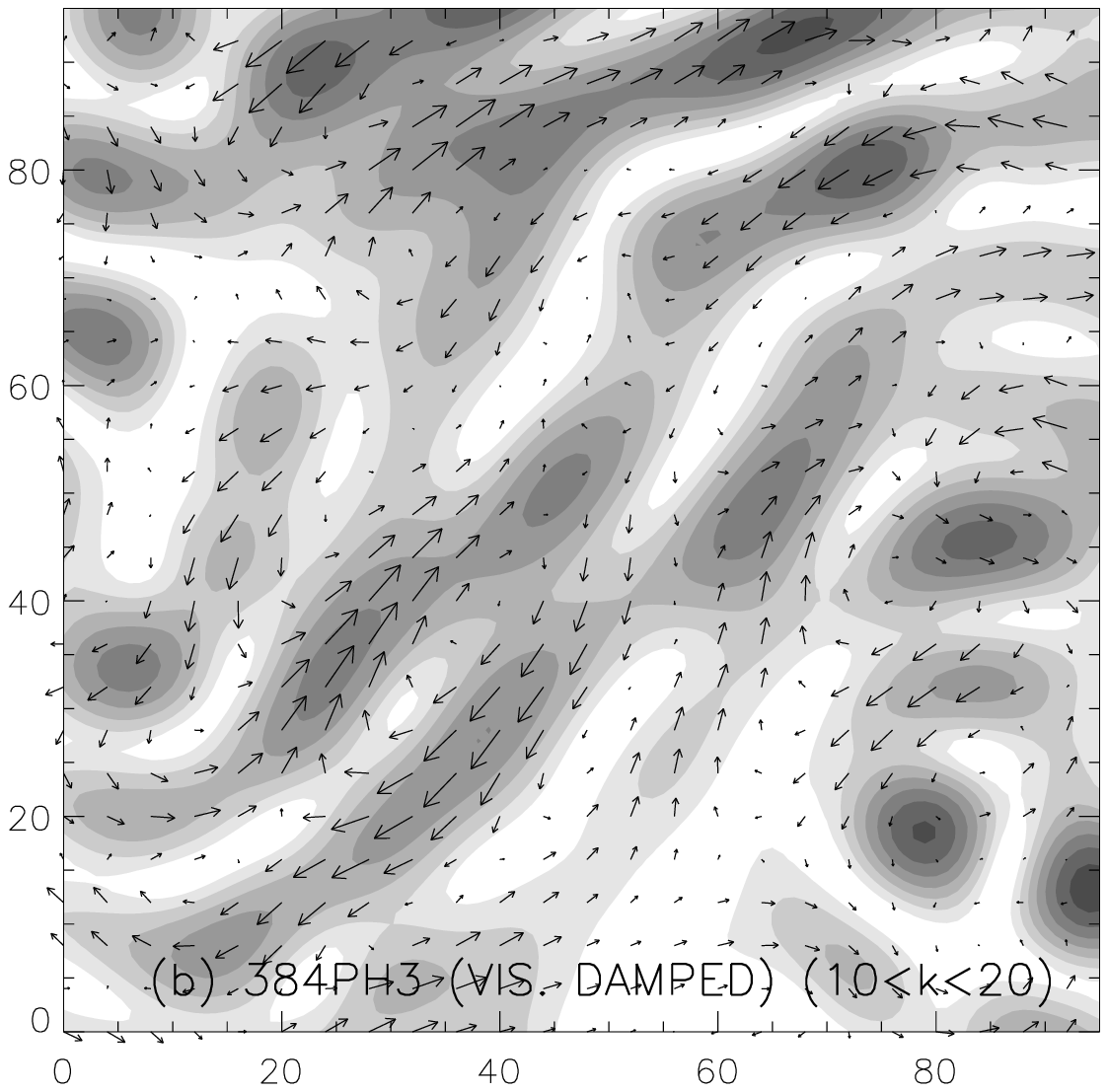}          \\
\includegraphics[width=0.49\textwidth]{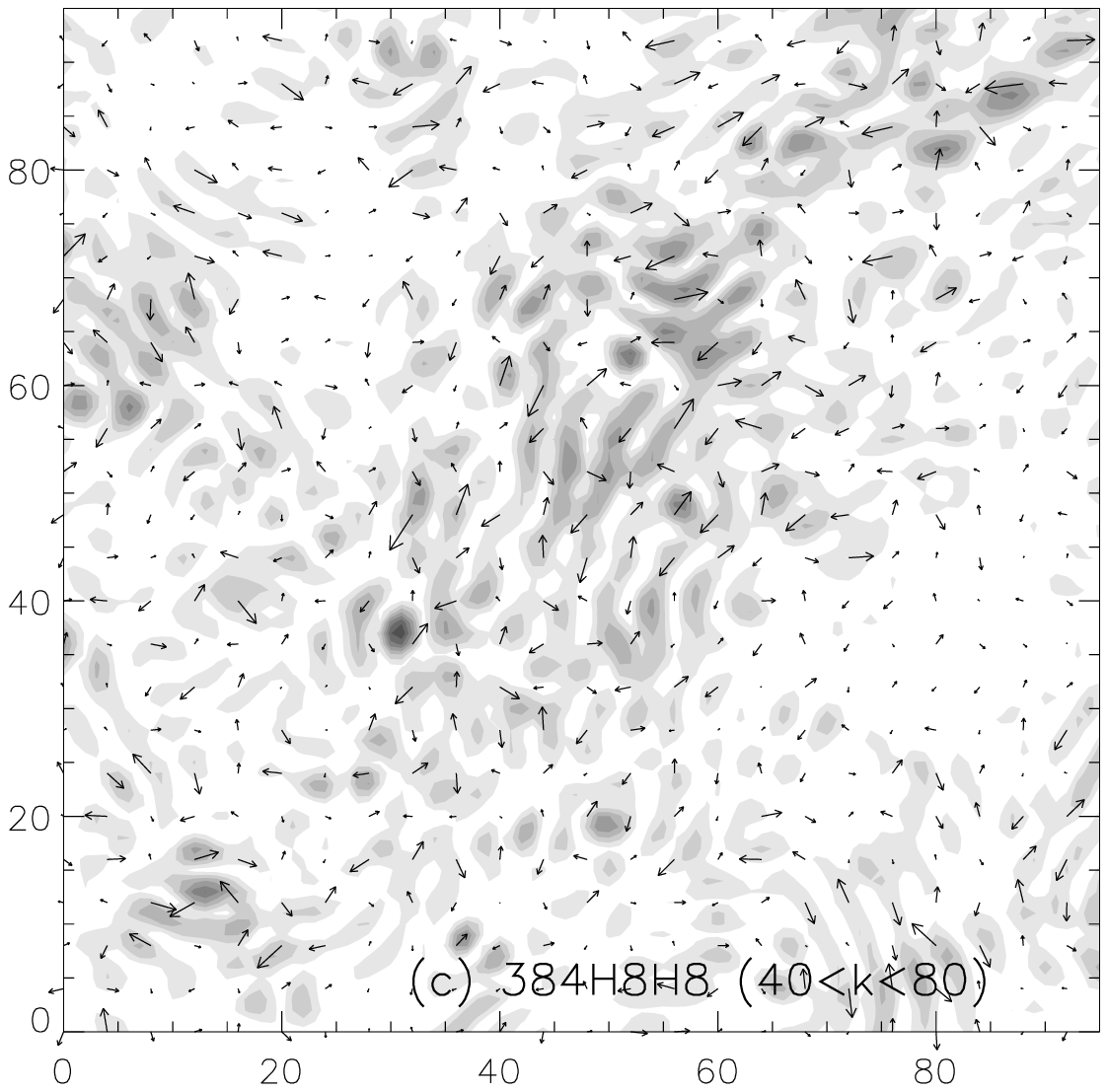}
\>
\includegraphics[width=0.49\textwidth]{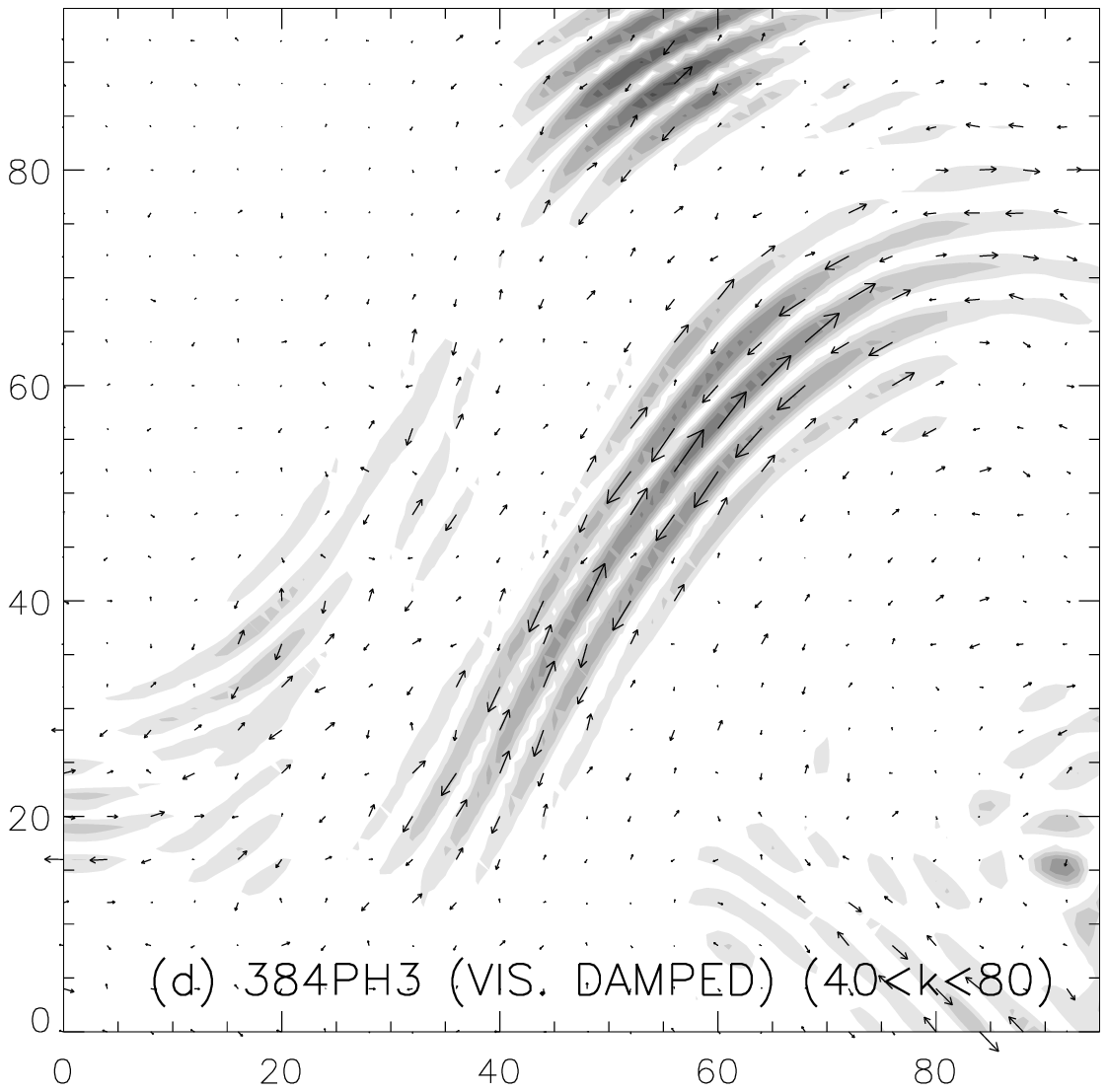} %  \\
\end{tabbing}
 \caption{ Magnetic structures in planes perpendicular to the mean 
   magnetic field.
  (a) and (c): Ordinary MHD turbulence. The distribution of magnetic energy
    is more or less smooth. We do not observe scale-dependency.
%%%------------------
%   (b) and (d): Viscosity-damped MHD turbulence.
%   The distribution of magnetic energy is very intermittent, and
%   is more pronounced at small scales.
   (b) and (d): Viscosity-damped MHD turbulence.
   In the Run 384PH3-B$_0$1, the viscosity-damped regime
   emerges below the dissipation scale at $k_c\sim 10$ (see Figure 1(b)).
   Panel (b) is {\it not} much different from panel (a) because
   the corresponding scale is not far from the dissipation scale.
   In panel (d), the distribution of magnetic energy is very intermittent.
   Panels (b) and (d) illustrate that
   intermittency is more pronounced at small scales.
%%%------------------
}
\label{fig_102040}
\end{figure*}

Figure \ref{fig_102040} shows magnetic structures in a plane perpendicular to
the mean field $\bf{B}_0$.
In  Figures \ref{fig_102040}(a) and (b), we plot
medium scale eddy shapes, using Fourier modes with $10<k<20$.
(Remember that in the viscosity-damped turbulence viscous damping occurs
around $k\sim 7$.)
In Figures \ref{fig_102040}(c) and (d), we plot
small scale eddy shapes using Fourier modes with $40<k<80$.
In ordinary turbulence (Figures  \ref{fig_102040}(a) and (c)), 
magnetic field structures are more or less smooth.
%%%----------------------
%In viscosity-damped turbulence (Figure  \ref{fig_102040}(b) and (d)), 
%we can see that
%magnetic field structures are highly intermittent.
 In viscosity-damped turbulence (Figure  3(b) and (d)),
    we can see that
    intermittency is more pronounced at small scales.
%%%----------------------
It is obvious that intermittency is scale-dependent in
viscosity-damped turbulence.

\begin{figure*}
  \includegraphics[width=0.49\textwidth]{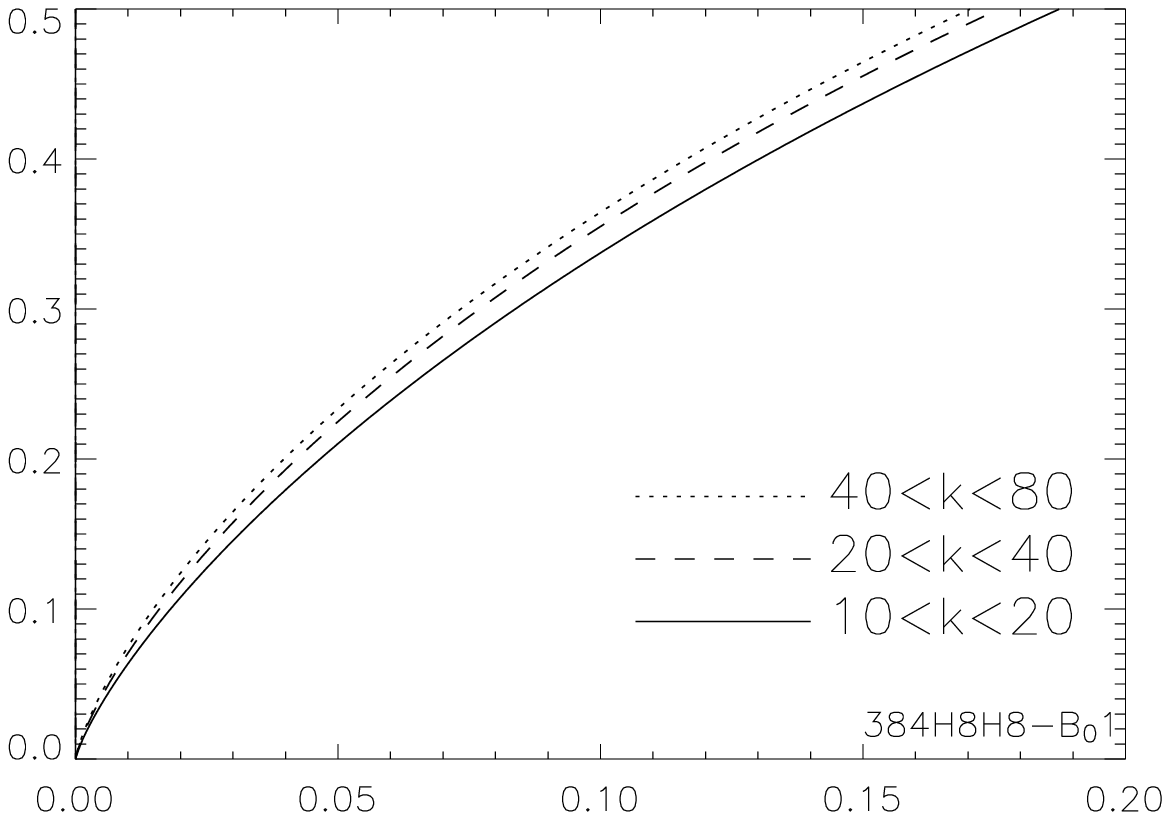}
\hfill
  \includegraphics[width=0.49\textwidth]{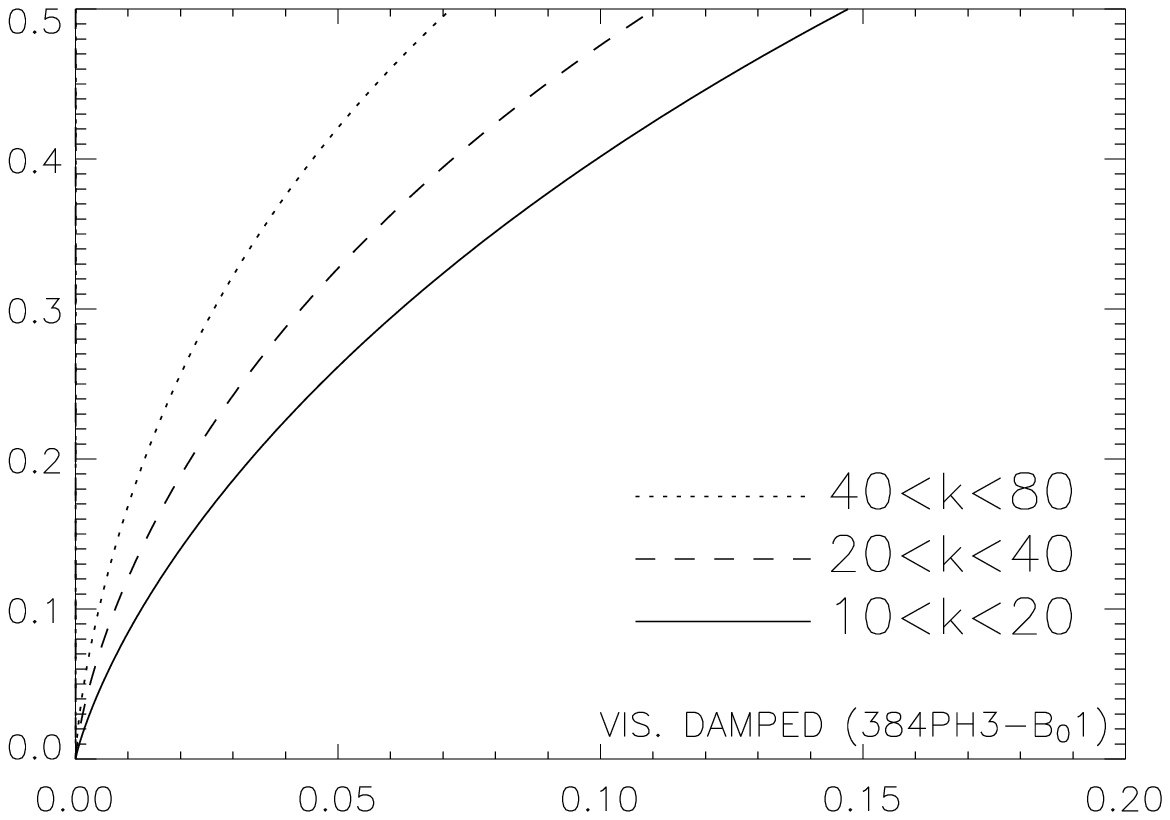}
  \caption{
      Fractional volume (X-axis) vs. 
      fractional magnetic energy in the volume (Y-axis)
    (a: {\it left}) Ordinary MHD strong turbulence:
        intermittency is independent of scale.
    (b: {\it right}) Viscosity-damped MHD turbulence:
        smaller scales show a higher concentration of magnetic energy.
}
\label{fig_fraction}
\end{figure*}

Figure \ref{fig_fraction} shows scale-dependent intermittency
more clearly.  On the x-axis we plot the volume fraction and on the
y-axis the fraction of the perturbed magnetic energy contained in that volume.
For example, in Figure \ref{fig_fraction}(b) we see that about 7\% of the
volume contains half the magnetic energy for modes with $40<k<80$.
However, at larger scales, 
the same energy occupies larger volume  (e.g. see the solid line).
In ordinary strong MHD turbulence (Figure \ref{fig_fraction}(a)),
there are only slight differences between the curves corresponding to 
different scales.

%%%%%%%%%%%%%%%%%%%%%%%%%%%%%%%%%%%%%%%%%%%%% conto
%%%%%%%%%%%%%%%%%%%%%%%%%%%%%%%%%%%%%%%%%%%%% conto

%%%%%%%%%%%%%%%%%%%%%%%%%%%%%%%%%%%%%%%%%%%%% fraction
%%%%%%%%%%%%%%%%%%%%%%%%%%%%%%%%%%%%%%%%%%%%% fraction

\section{High order statistics}
Spectra do not provide a full description of
turbulence.  Another useful source of information is the scaling of 
high order structure functions.
The p-th order (longitudinal) velocity structure function $SF_p$ 
and scaling exponents
$\zeta(p)$ are defined as
\begin{equation}
SF_p({\bf r}) \equiv \langle 
  | \left[ {\bf v}({\bf x}+{\bf r})-{\bf v}({\bf x}) \right] \cdot {\bf \hat r} |^p
  \rangle \propto r^{\zeta (p)},
\label{structure}
\end{equation}
where the angle brackets denote averaging over {\bf x}.
In this section, we discuss the scaling of $\zeta (p)$ in ordinary and 
viscosity-damped turbulence.

\subsection{Ordinary turbulence}

The 
scaling relation suggested by She \& Leveque (1994) 
contains
three parameters (see Politano \& Pouquet 1995; M\"{u}ller \& Biskamp 2000): 
$g$ is related to the scaling $v_l\sim l^{1/g}$, $x$ related to
the energy cascade rate $t_l^{-1}\sim l^{-x}$, and C, the co-dimension of the
dissipative structures:
\begin{equation}
\zeta(p)={p\over g}(1-x)+C\left(1-(1-x/C)^{p/g}\right).
\label{She-Leveque}
\end{equation}
For incompressible hydrodynamic turbulence, She \& Leveque (1994) obtained
\begin{equation}
\zeta^{SL}(p)=p/9+2 \left[1-(2/3)^{p/3} \right],
\label{She-Leveque1}
\end{equation}
using $g=3$, $x=2/3$, and $C=2$, implying that
dissipation happens over 1D structures (e.g. vortices).

In 3-dimensional MHD turbulence, there have been three recent developments
for the scaling of the structure function exponents $\zeta(p)$.
First, M\"{u}ller \& Biskamp (2000) extended the She-Leveque model to
incompressible MHD turbulence and obtained
\begin{equation}
\zeta^{MB}(p)=p/9+1-(1/3)^{p/3},
\label{eq_MulB00}
\end{equation}
where they used $g=3$, $x=2/3$, and assumed that the 
dissipating structures
are 2-dimensional (i.e $C=3-2=1$).
Their numerical calculations for decaying MHD turbulence without mean field
show that 
scaling exponents for ${\bf z}^{\pm}\equiv {\bf v} \pm {\bf b}$
follow this relation very closely.
Second, CLV02a studied MHD turbulence in a 
strongly magnetized medium and showed
that turbulent motions (i.e. {\it velocity}) in planes perpendicular to
the local mean field directions follow the original She-Leveque scaling
(equation (\ref{She-Leveque1})). Physically this might mean that the
dissipative structures are again 1D hydrodynamic-type vortices. 
However, more probably, this may just mean that the dissipation structures 
look like one-dimensional in
the (perpendicular) slices 
of 3D MHD turbulence.
Third, Boldyrev (2002) assumed that in compressible turbulence dissipation
happens over 2D structures, i.e. $C=1$ and eq.~(\ref{eq_MulB00}) for
$\zeta$. This scaling was supported by
compressible MHD simulations (Padoan et al. 2003a).
Padoan et al. (2003a) showed that the velocity field in highly super-sonic 
MHD turbulence follows the scaling given by eq.~(\ref{eq_MulB00})
scaling, while
in subsonic MHD turbulence it follows the scaling in equation (\ref{She-Leveque1}).

Combined together these results look puzzling.
One can see easily that there are at least two apparent contradictions.
First, Cho et al.'s (CLV02a) result conflicts with M\"{u}ller \& Biskamp's result.
This discrepancy
could be attributed to the different simulation parameters.
For example, CLV02a used a strong mean field while MB00 used no mean field.
However, according to M\"{u}ller \& Biskamp (2003),
the difference in $B_0$ does not resolve this problem.
Second, Padoan et al. (2003a) is at odds with M\"{u}ller \& Biskamp (2000):
Padoan et al.'s  results for low Mach number 
do not converge to the incompressible case. 
The difference in $B_0$ looks
marginal: M\"{u}ller \& Biskamp (2000) used a zero mean magnetic field 
while Padoan et al. (2003a) used a weak mean magnetic field.

\begin{figure*}
\begin{tabbing}
  \includegraphics[width=0.49\textwidth]{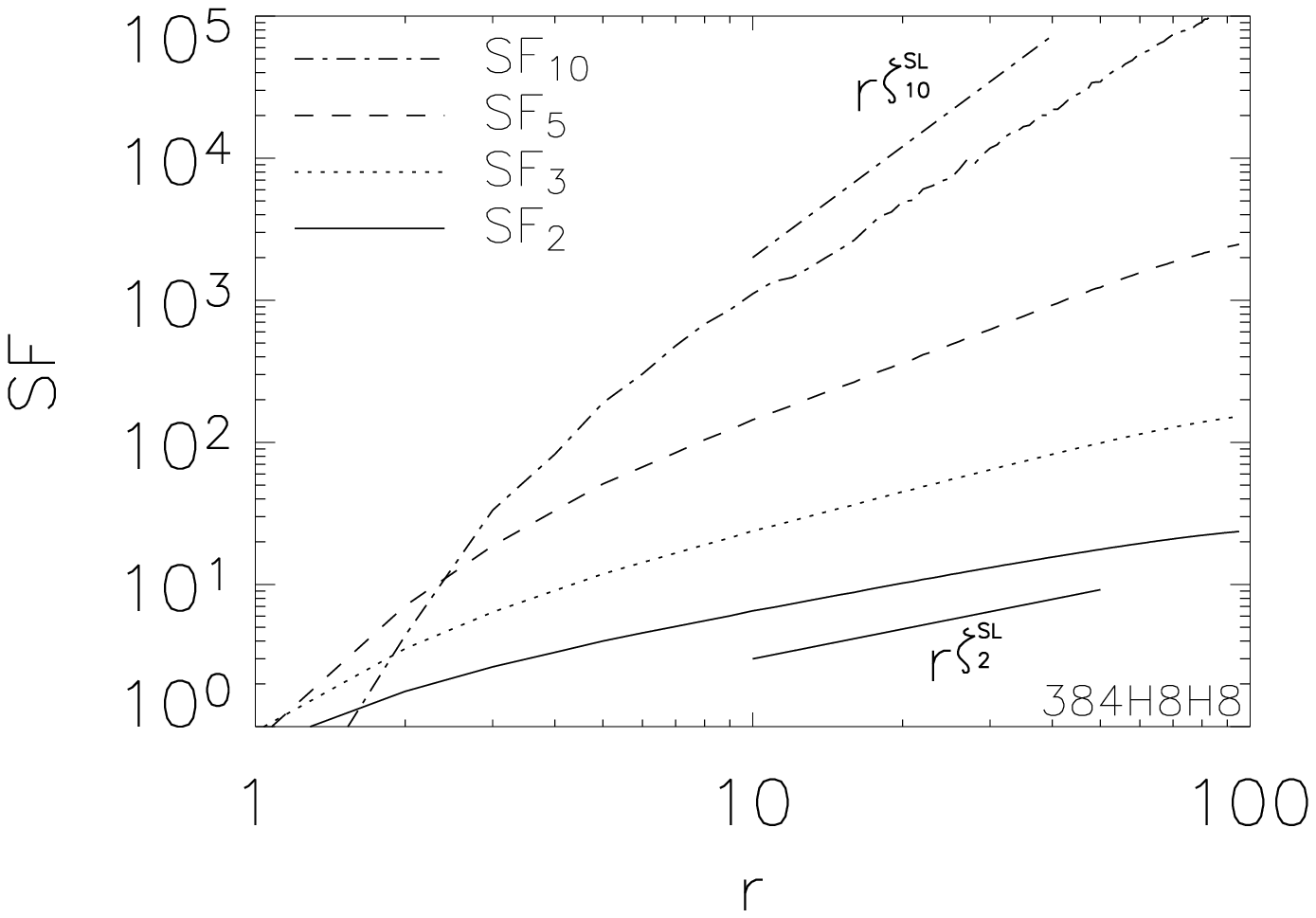}
\=
 \includegraphics[width=0.49\textwidth]{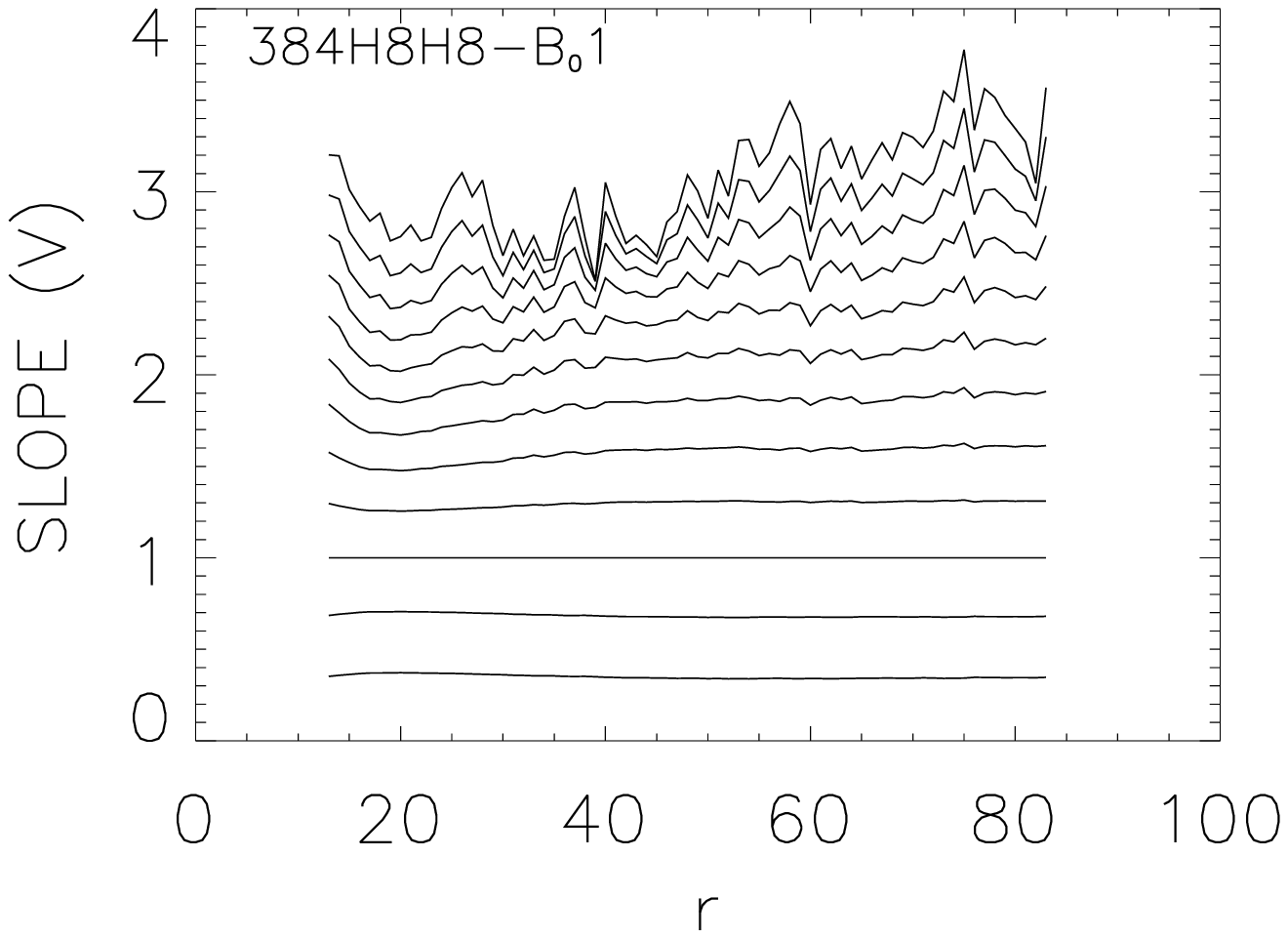}   \\
   ~~~~~~~~~~~~~~~~~~~~~~~~~~~~~~~~~~~~~~~~~~~(a) 
\> ~~~~~~~~~~~~~~~~~~~~~~~~~~~~~~~~~~~~~~~~~(b) 
\\  \\
  \includegraphics[width=0.49\textwidth]{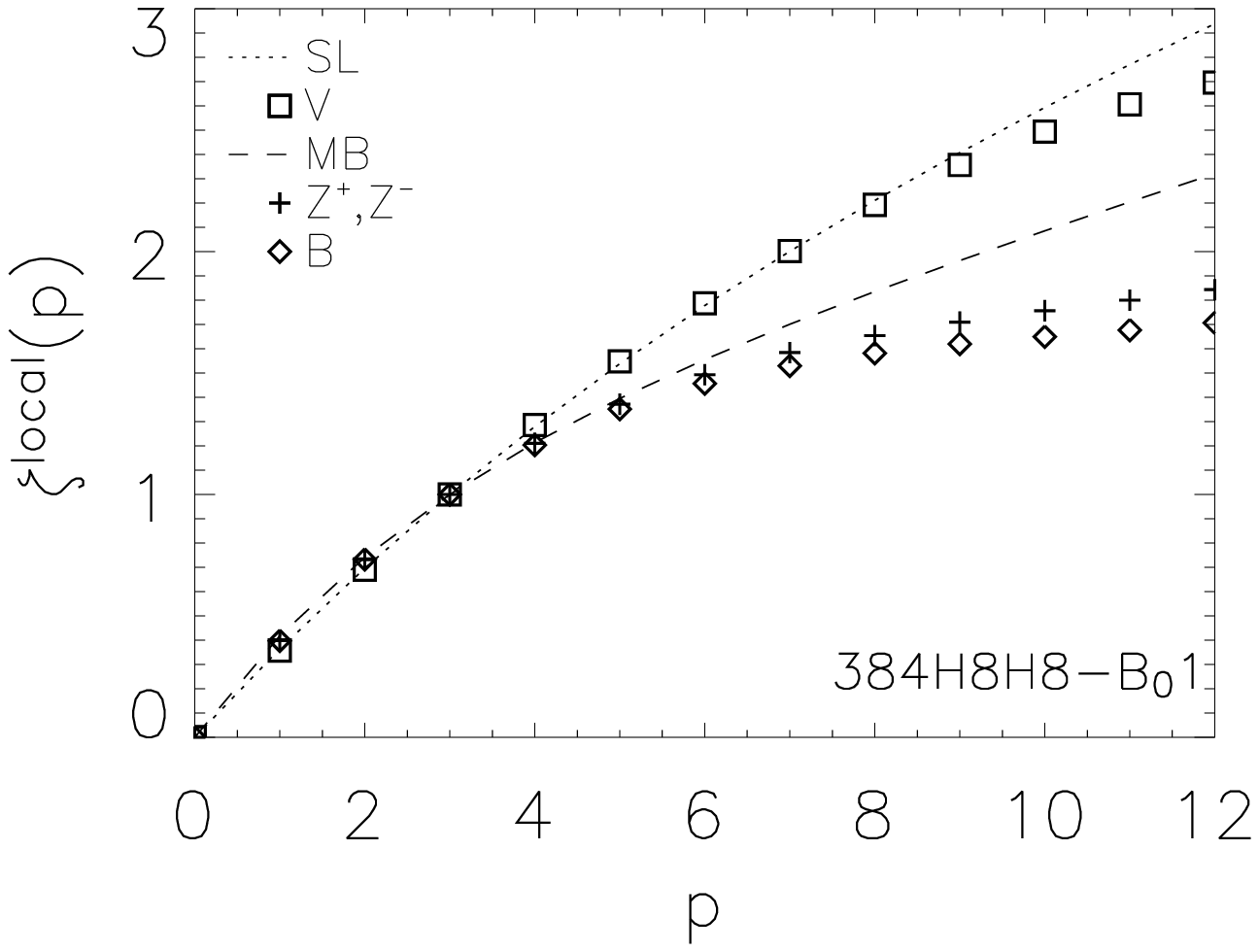}
\>
 \includegraphics[width=0.49\textwidth]{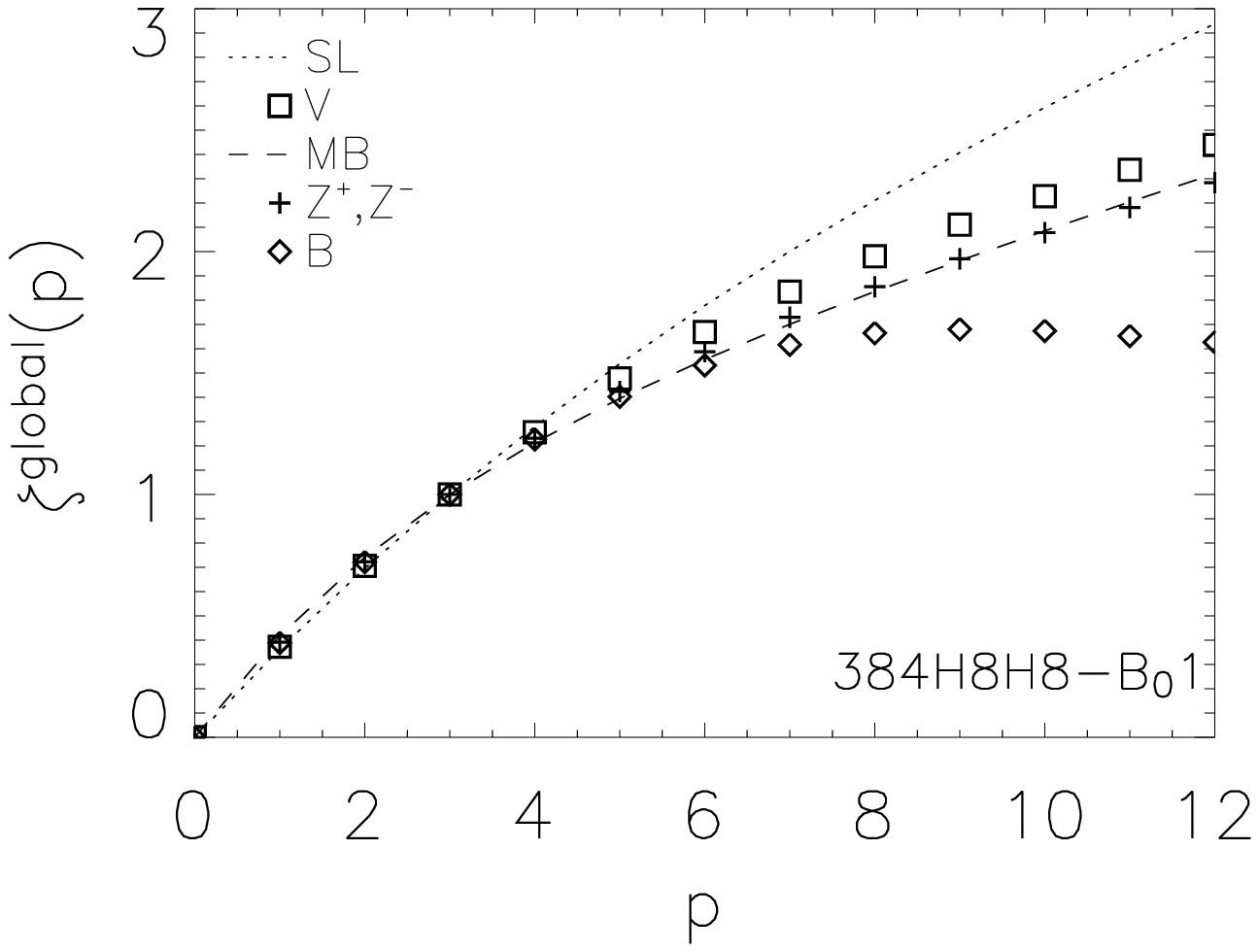} \\
   ~~~~~~~~~~~~~~~~~~~~~~~~~~~~~~~~~~~~~~~~~~~(c) 
\> ~~~~~~~~~~~~~~~~~~~~~~~~~~~~~~~~~~~~~~~~~(d) 
\end{tabbing}
  \caption{
      The intermittency of ordinary turbulence (384H8H8-B$_0$1).
    (a: {\it upper-left}) Velocity structure functions 
       (multiplied by arbitrary constants)
        in planes
        perpendicular to the local mean magnetic fields.
    (b: {\it upper-right}) Differential slopes normalized 
       by that of the third-order structure
        function: $ [d \ln SF_p/d \ln r]/
                    [d \ln SF_3/d \ln r]$, where $SF_p$ is the p-th order
        longitudinal velocity structure function calculated in planes
        perpendicular to the local mean magnetic fields.
        The slope for $SF_1$ is the bottom curve and that for $SF_{12}$ 
        is the top curve.
    (c: {\it lower-left}) Normalized structure function 
        exponents in perpendicular directions
        in the local frame. 
        The velocity exponents show a scaling similar to the She-Leveque model.
        The magnetic field shows a different scaling.
    (d: {\it lower-right}) Normalized structure function 
        exponents in the global frame.
        Note that the result for $z^{\pm}$ is very similar to 
        the M\"{u}ller-Biskamp model.
}
\label{fig_int_ord}
\end{figure*}

These contradictions have led us to revisit the issue of intermittency
in strong MHD turbulence.
In Figure \ref{fig_int_ord}(a), we plot longitudinal velocity 
structure functions $SF_p$.
The calculations are done in the {\it perpendicular} planes 
in the local frame of reference, where
the {\it parallel} axis is aligned with the local mean magnetic field
(see CLV02a):
\begin{equation}
SF_p^{local}({\bf r}) \equiv \langle 
  | \left[ {\bf v}({\bf x}+{\bf r}_{\perp})-
           {\bf v}({\bf x}) \right] \cdot {\bf \hat r}_{\perp} |^p
  \rangle,
\end{equation} 
where ${\bf r}_{\perp}$ is perpendicular to local mean field and ${\bf \hat r}_{\perp}$
is the unit vector parallel to ${\bf r}_{\perp}$.
We show the second, third, fifth, and 10th order structure functions.
We  observe that
the slopes are slightly shallower than those of She-Leveque model.
Here the results are for longitudinal structure functions, which 
mostly reflect Alfven mode statistics in the 
perpendicular planes.  However, when we use normal structure functions 
($ \langle 
  | \left[ {\bf v}({\bf x}+{\bf r}_{\perp})-
           {\bf v}({\bf x}) \right] |^p
  \rangle $), which reflect both Alfven and pseudo-Alfven statistics,
the $\zeta (p)$'s show a larger deviation from the She-Leveque model.
This may be understood if the pseudo-Alfven modes have different
scaling exponents.
We can separate out the scaling of pseudo-Alfven modes when we use 
transverse structure functions in the perpendicular planes in
local frame.
We can also capture them using 
longitudinal structure functions in the parallel directions in local frame.
However, we will not pursue this issue further here.

In Figure \ref{fig_int_ord}(b), we plot the normalized 
differential slope, $ [d \ln SF_p/d \ln r]/
                      [d \ln SF_3/d \ln r]$,
for local frame velocity structure functions.
In the actual calculations, we use
$ [d \ln SF_p/d \ln r](r) \approx 
    \ln [SF_p(r+4) / SF_p (r-4)]/ \ln [(r+4)/(r-4)]$.
{}From the Figure, we obtain the $\zeta (p)$'s by averaging the differential slope
over $r \in [20, 45]$.
In general, normalizing the differential slope using $SF_3$
gives better-defined 
scaling exponents, which was noted in M\"{u}ller \& Biskamp (2003) and by
Boldyrev (private communication).

In Figure \ref{fig_int_ord}(c), we plot the $\zeta (p)$'s obtained this way for
velocity, magnetic field, and $z^{\pm}$.
The scaling exponents for velocity show reasonable agreement with
the She-Leveque model, in agreement with CLV02a. 
Those for the magnetic field follow the M\"{u}ller-Biskamp model for small $p$
but level off and do not change much when $p>7$.
Those for Elsasser variables  $z^{\pm}$ lie between
those for the velocity and magnetic fields.

In Figure \ref{fig_int_ord}(d), we plot $\zeta (p)$'s
calculated in the global frame, in which coordinate axes are aligned with
the usual Cartesian axes.
The method we used for  Figure \ref{fig_int_ord}(d) is
similar to the methods used by 
M\"{u}ller \& Biskamp (2000) and Padoan et al. (2003a), apart from
the fact that our turbulence is strongly magnetized.
It is worth noting that we get a good agreement with the
M\"{u}ller-Biskamp (2000) model for $z^{\pm}$ variables.

Our first result is that the scaling exponents obey the rule
$\zeta^{magnetic} (p) < \zeta^{z}(p)< \zeta^{velocity}(p)$. 
It matters whether one uses Elsasser variables, velocity or magnetic
field to determine the dimension of the dissipation structures. Our
second result is that the dimension of the dissipation structures
looks different when viewed in local and global frames of reference\footnote{
  Note that the calculation is done 
  for the perpendicular planes in the local frame, while
  it is done for all directions in the global frame.
}.
It looks as if one dimensional vortices merge into two-dimensional
sheets when viewed from the global system of reference. These differences
should make us wary of a naive association of the corresponding
parameters in eq.(\ref{She-Leveque}) with the dimensions of the dissipation
structures in MHD turbulence. 

This result can eliminate the contradiction between
M\"{u}ller \& Biskamp (2000) and CLV02a:
the former used the Elsasser variables in the global frame 
and the latter used velocity in the local frame
for studying scaling exponents (see Figure \ref{fig_int_ord}(c) and (d)).
Similarly, the conflict between M\"{u}ller \& Biskamp (2000) and
Padoan et al. (2003a) can be partially relieved:
the latter used velocity and, therefore, we
expect that the scaling exponents in the latter are larger than
those in the former.
However, it is not certain from this that 
the discrepancy between the Padoan et al. (2003a) scalings at low
Mach number and those in M\"{u}ller \& Biskamp (2000) in
the incompressible case has been completely resolved.
It is also unclear why the magnetic and velocity fields have 
different scalings. The sense of the difference suggests that the magnetic
field is significantly more intermittent than the velocity field. 
Apparently this topic deserves further theoretical research. 

%%%%%%%%%%%%%%%%%%%%%%%%%%%%%%%%%%%%%%%%%%%%% Intermittency: ordinary
%%%%%%%%%%%%%%%%%%%%%%%%%%%%%%%%%%%%%%%%%%%%% Intermittency: ordinary

\subsection{ High order structure functions in decaying compressible
MHD turbulence}

Above we speculated that there may not be a contradiction between the
results by M\"{u}ller \& Biskamp (2000) and Padoan et al. (2003a). 
To test this we performed a numerical simulation for 
decaying compressible MHD turbulence without mean magnetic field.
The compressible MHD code is described in detail in Cho \& Lazarian (2002).   
We use the same numerical scheme described above.
The initial kinetic and magnetic energy spectra are
\begin{equation}
  E_v(k) \approx E_b(k) \approx k^2e^{-k^2/k_0^2},  \label{eq_init_sp}
\end{equation}
where we take $k_0=2$.
The initial phases of the individual Fourier components are random.
The initial sonic Mach number (the ratio of the rms velocity to the sound speed)
is $\sim 0.7$ and the density is constant.

\begin{figure*}
  \includegraphics[width=0.32\textwidth]{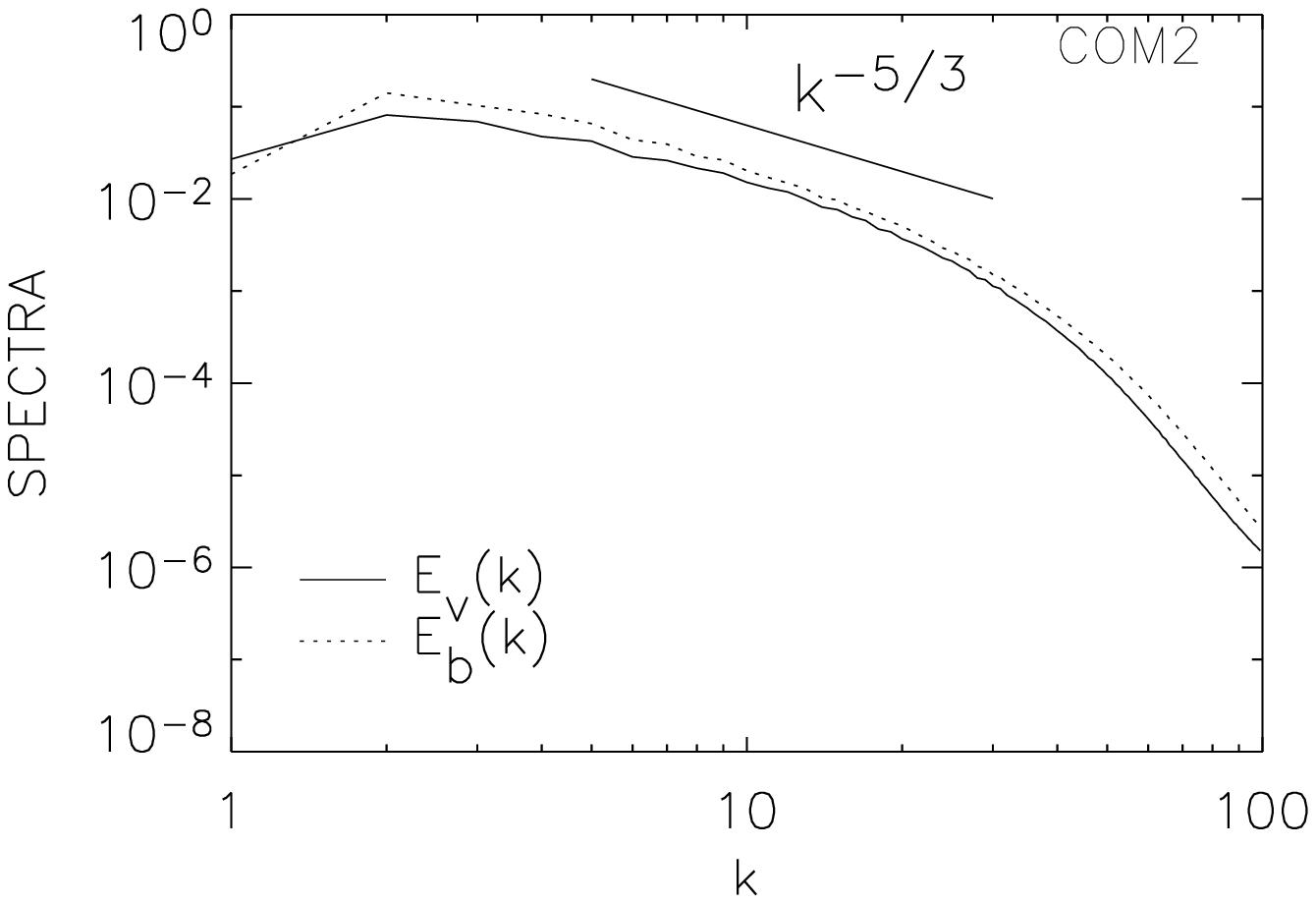}
\hfill
  \includegraphics[width=0.32\textwidth]{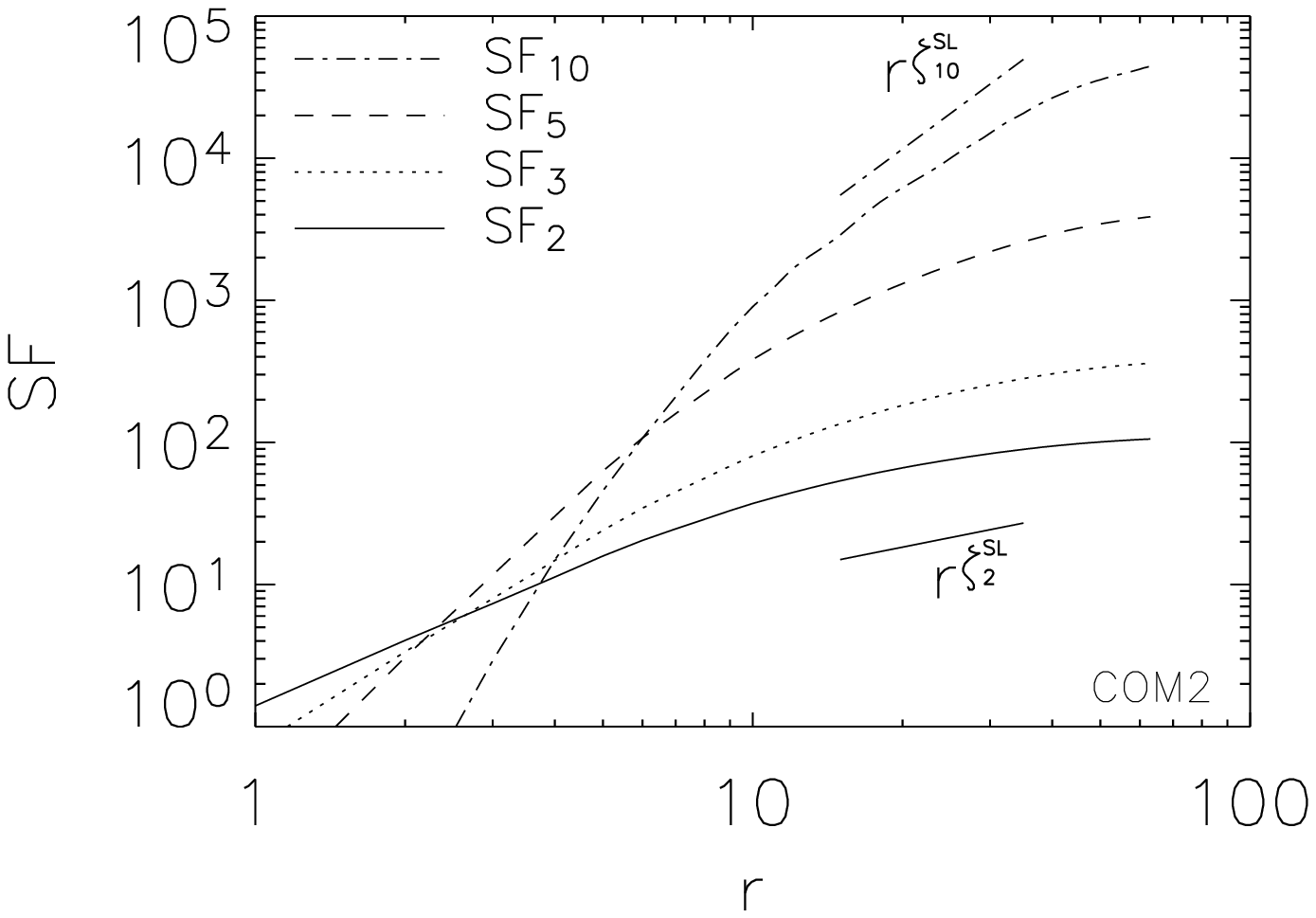}
\hfill
  \includegraphics[width=0.32\textwidth]{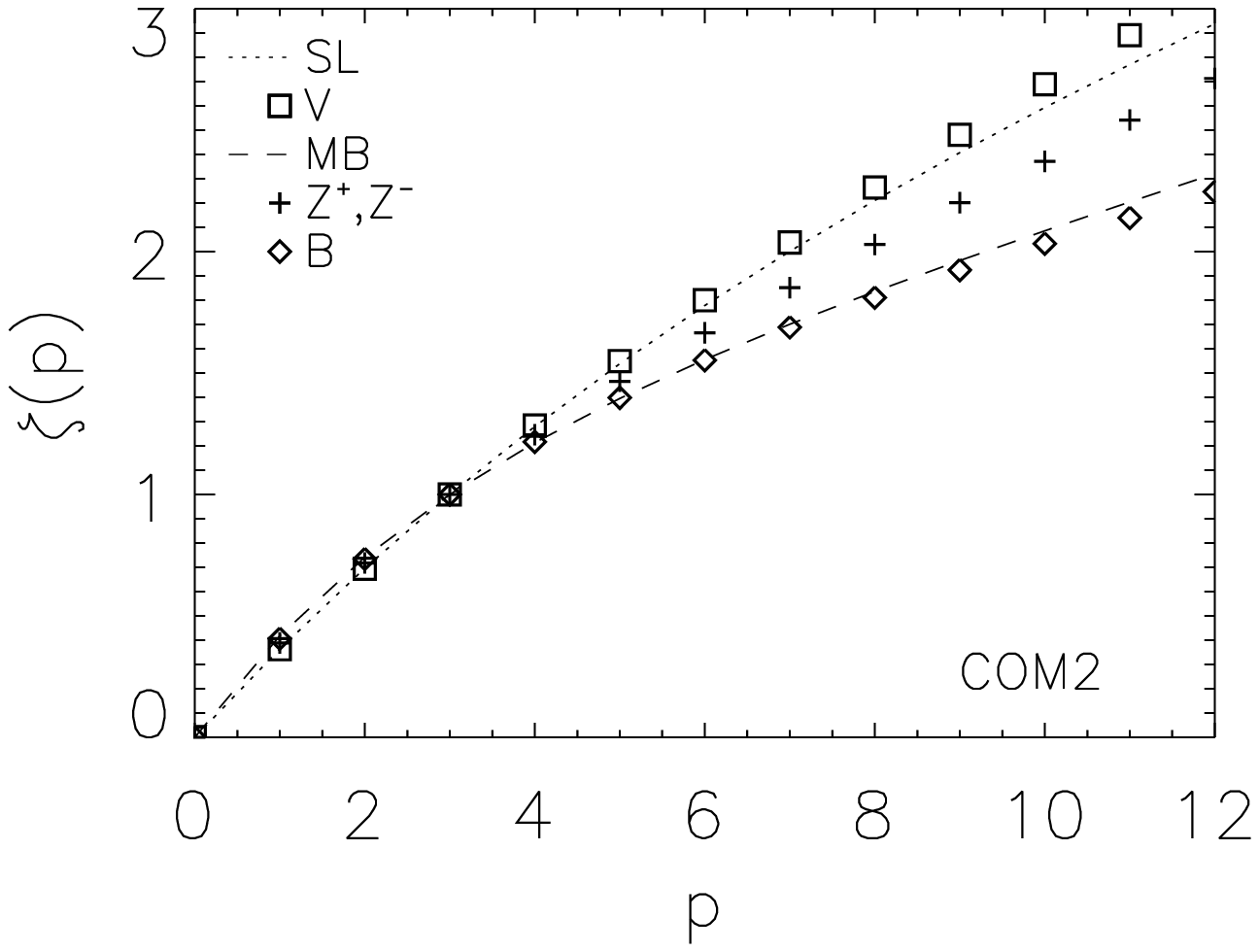}
  \caption{
      Compressible decaying MHD turbulence (COM2).
      The mean magnetic field is zero: B$_0$=0.
    (a: {\it left}) Spectra.
    (b: {\it middle}) Structure functions in the global frame.
    (c: {\it right}) Normalized scaling exponents 
        $\zeta (p)$ in the global frame.
        The magnetic and velocity fields show different scalings.
}
\label{fig_compB0}
\end{figure*}

Initially only small-$k$ Fourier components near $k=k_0$ are excited,
but as the energy cascade begins to operate the 
large-$k$ Fourier components are excited.
We measure high-order structure function statistics after
the energy spectra develop inertial range
%%%WHAT DOES THIS MEAN EXACTLY?
(Figure \ref{fig_compB0}(a)). 
Initially the velocity and magnetic fields have almost identical
spectra (equation (\ref{eq_init_sp})).
It is interesting that the magnetic energy spectrum is larger than
its kinetic counterpart at later times (see Biskamp \& M\"{u}ller 2000 for
a similar result in the incompressible case).
Figure \ref{fig_compB0}(b) shows the raw velocity structure functions.
Figure \ref{fig_compB0}(c) shows the normalized $\zeta (p)$ 
(i.e.$\zeta (p)/\zeta (3)$), which we obtain
by averaging the slopes between $r=17$ and $r=34$.
Interestingly, it appears that the velocity closely follows She-Leveque scaling 
while the 
magnetic field follows M\"{u}ller-Biskamp scaling.
The velocity scaling we find is indeed consistent with the result in
Padoan et al. (2003a).
However, since we do not resolve a sufficiently long inertial range (see the spectra),
the results in Figure \ref{fig_compB0}(c) should be regarded as very tentative.
We will present a comprehensive study of higher order statistics elsewhere.

%%%%%%%%%%%%%%%%%%%%%%%%%%%%%%%%%%%%%%%%%%%%%Fig begin
%%%%%%%%%%%%%%%%%%%%%%%%%%%%%%%%%%%%%%%%%%%%%end Fig

\subsection{Viscosity-damped turbulence}

\begin{figure*}
 \includegraphics[width=0.32\textwidth]{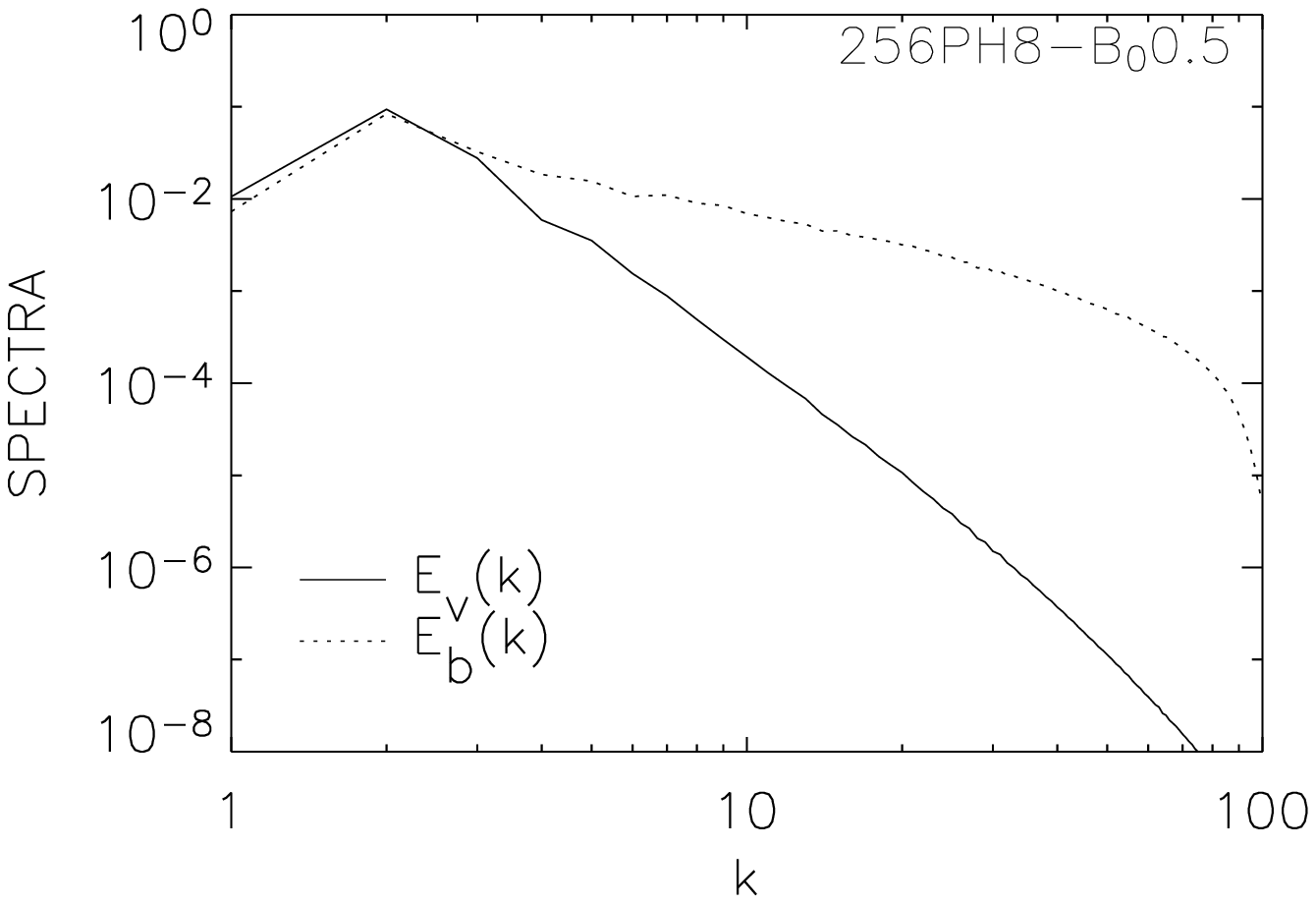}
\hfill
 \includegraphics[width=0.32\textwidth]{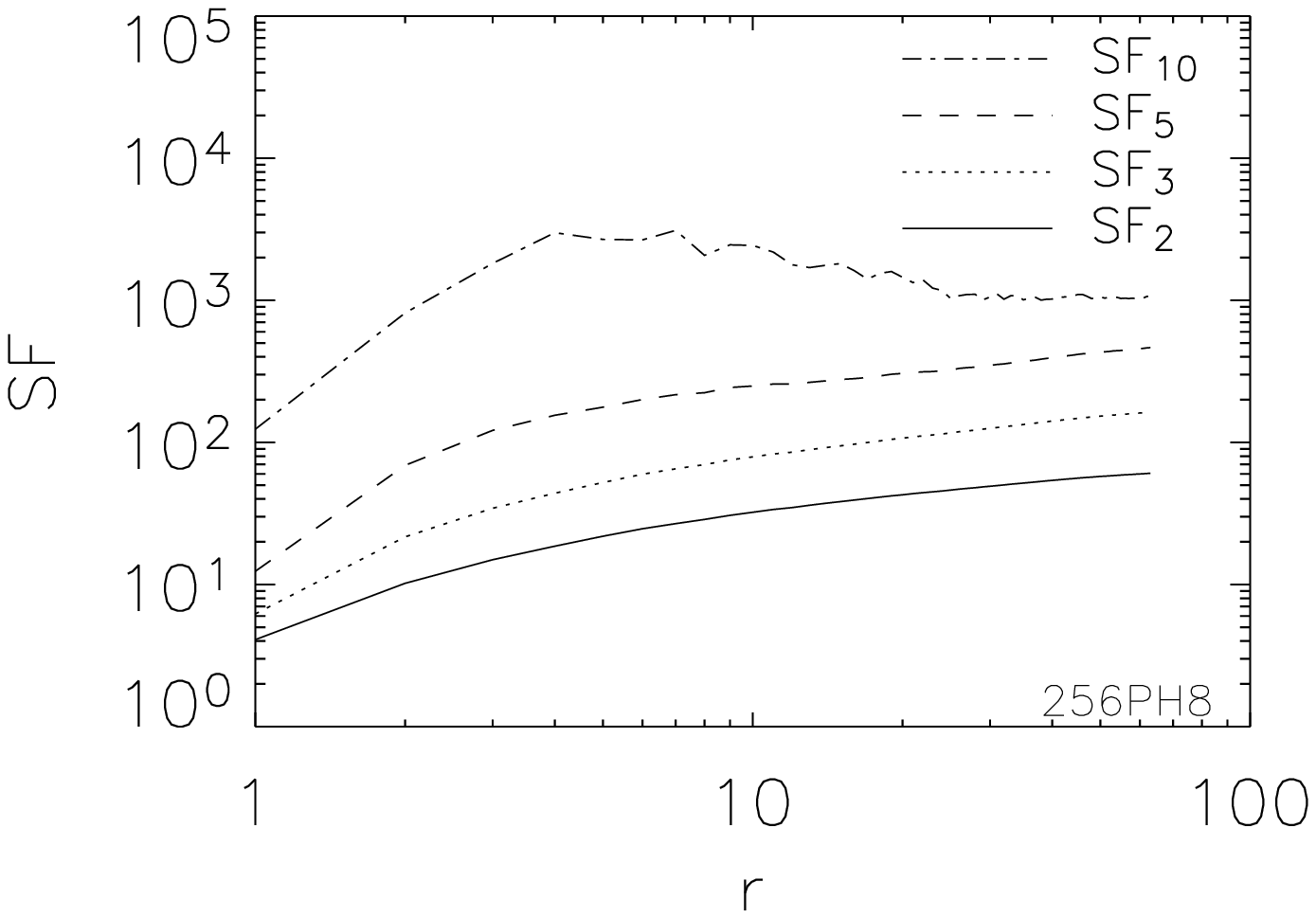}
\hfill
 \includegraphics[width=0.32\textwidth]{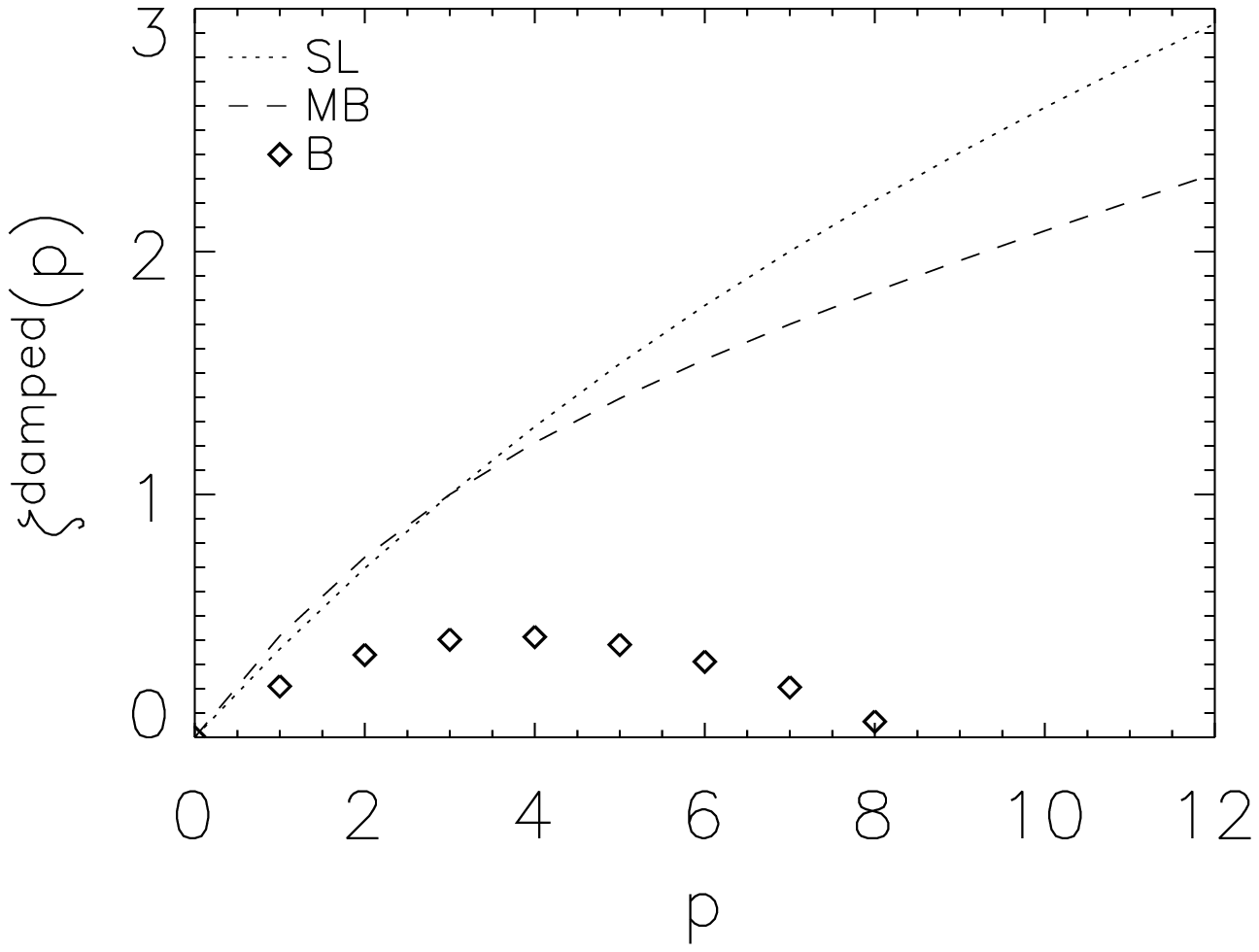}
 \caption{
   Viscous damped turbulence (256PH8-B$_0$0.5).
   (a: {\it left}) Spectra.
   (b: {\it middle}) Magnetic structure functions in the local frame.
       (Those calculated in the global frame show similar behaviors.)
   (c: {\it right}) Magnetic structure function exponents, 
       $\zeta (p)$, in the local frame
       (not normalized).
       The observed scaling exponents are at least close to
       the expected asymptote $\zeta (p)=0$.
}
\label{fig_raw_vis}
\end{figure*}

The run 384PH3-B$_0$1 has a very limited inertial range for the magnetic field,
and is not suitable for the study of scaling exponents.
We use the run 256PH8-B$_0$0.5 ($\nu=0.06$ and $\eta=hyper-diffusion$) instead.
In this run, the mean field strength is reduced to 0.5 because
the rms velocity is $\sim$0.4 due to strong viscous damping.
The energy spectra is plotted in Figure \ref{fig_raw_vis}.
This run also clearly shows scale dependent intermittency.
The parallel wavenumber $k_{\|}$ shows a very slow increase,
a factor of $\sim 2$ from
$k=5$ to $k=70$.
Viscous damping occurs right at the energy injection scale
(Figure \ref{fig_raw_vis}(a)).
The magnetic spectrum shows a slope roughly consistent with $-1$.
Here we consider only the magnetic scaling exponents.

%%%%%%%%  Modification begins here.  Apr 11, 2003 %%%%%%

Figure \ref{fig_raw_vis}(b) shows the structure functions for magnetic field.
The calculation is done in perpendicular planes in local frame.
Figure \ref{fig_raw_vis}(c) shows that
the scaling exponents becomes negative when $p$ is larger than 8.

In the original She-Leveque model, they first obtained the scaling
relation for high order statistics of
energy dissipation averaged over a ball of size $l$,
$\epsilon_l$:
$<\epsilon_l^p>\sim l^{\tau_p}$ with $\tau_p^{SL}=-2p/3+2[1-(2/3)^p]$.
Then, using the Kolmogorov refined similarity hypothesis,
$\epsilon_l \approx v_l^3/l$ or $v_l\approx l^{1/3}\epsilon_l^{1/3}$,
they obtained
\begin{equation}
\zeta^{SL} (p) = p/3 + \tau_{p/3}.
\end{equation}
Note that $\tau_{p}<0$ and, hence, $\zeta^{SL} (p) < p/3$.
In viscosity-damped MHD turbulence, we cannot use
the Kolmogorov refined similarity hypothesis directly.
Instead, from $\epsilon_l \approx b_l^2/l^0$ or $b_l\approx\epsilon^{1/2}$,
we expect
\begin{equation}
\zeta^{damped} (p) = \tau_{p/2}.    \label{eq_zeta_damped}
\end{equation}

{} As in She-Leveque (1994; see also Politano \& Pouquet 1995), we may
write
\begin{equation}
\tau_p = -xp + C[1-(1-x/C)^p],
\end{equation}
where $x$ and $C$ are defined as in
equation (\ref{She-Leveque}).
When we substitute $x=0$, we get $\tau_{p}=0$ and $\zeta^{damped} (p)=0$.
However, the 2-parameter She-Leveque model in equation (\ref{She-Leveque})
may not be applicable to the case of $x=0$.

Despite the evident problems with extending the She-Leveque model
into the viscosity-damped regime, 
a crude estimate for the $\zeta(p)$ can be obtained from the theoretical
model discussed  in \S 2.  Given the extreme intermittency of the magnetic
field we expect that equation (\ref{structure}) will give
\be
SF_p(l) \sim \hat b_l^p\phi_l \sim l^{1-p/2},
\ee
or
\be
\zeta(p)=1-{p\over2}.
\ee
However, the rise in magnetic field strength at small length scales introduces
a competing contribution from the resistive scale, since for $p>2$ the rise
in local magnetic field strength overwhelms the decrease in the filling factor as
a function of scale.  This small scale contribution is not a function of $l$
and when it dominates we expect $\zeta(p)\sim 0$.  Putting all this together, we
see that our simple model for the viscosity-damped regime predicts that $\zeta(1)$
will lie between $0$ and $0.5$, $\zeta(2)\sim 0$, and when $p>2$ we expect that
$\zeta(p)$ will become negative and then asymptote to $0$ at large $p$.

Comparing this prediction to Figure (\ref{fig_raw_vis}) we see that this naive
prediction enjoys some qualitative success, but is certainly not exactly correct.
Part of the
problem may lie in the limited resolution of our numerical simulation.  For
example, the expectation that the magnetic energy spectrum follows a $k^{-1}$
scaling is fairly compelling, but in that case we would expect $\zeta(2)=0$.
The scaling in the simulation is slightly steeper (Figure \ref{fig_raw_vis}(a)).
The failure of the higher order exponents to match the theoretical model
may be part of the same problem, but it is also likely to reflect the crude
nature of the model.  Kolmogorov turbulence theory gives a passable match
to $\zeta(2)$, but fails noticeably at higher $p$, and our model for the
viscosity-damped MHD regime is not particularly more sophisticated.  
 
%%%%%%    Modification ends here %%%%%%%%%%%%%%

%%%%%%%%%%%%%%%%%%%%%%%%%%%%%%%%%%%%%%%%%%%%%Fig begin
%%%%%%%%%%%%%%%%%%%%%%%%%%%%%%%%%%%%%%%%%%%%%end Fig

\section{Testing and Discussion}

\subsection{ Viscosity-damped turbulence with physical diffusion}

How real are the structures that we observe in the new regime of MHD turbulence?
Could they be a numerical artifact?

We have used a third or higher order hyper-diffusion term to minimize
the effects of  
magnetic diffusion.
However, in general, high order hyper-diffusion suffers from
a bottle-neck effect, which is characterized by a flattening 
of the energy spectrum
near the dissipation scale.
Therefore it is necessary to check 
if the tail of the magnetic fluctuations is a real physical effect
and not due to a bottle-neck.
Here, we present a run with a physical magnetic diffusion term
(Run 256PP-B$_0$1; $\nu=0.015$, $\eta=0.001$).

\begin{figure*}
  \includegraphics[width=0.49\textwidth]{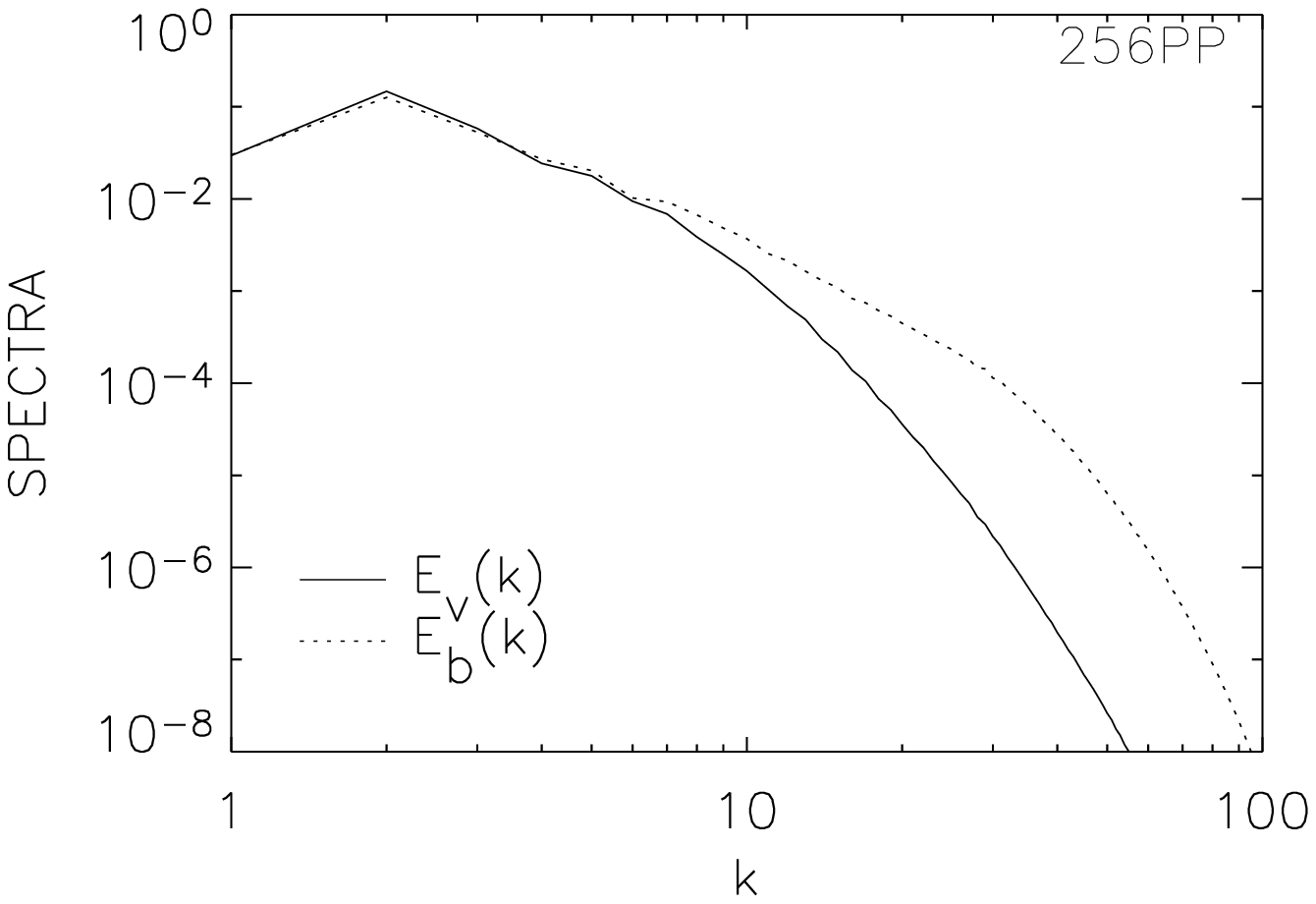}
\hfill
  \includegraphics[width=0.49\textwidth]{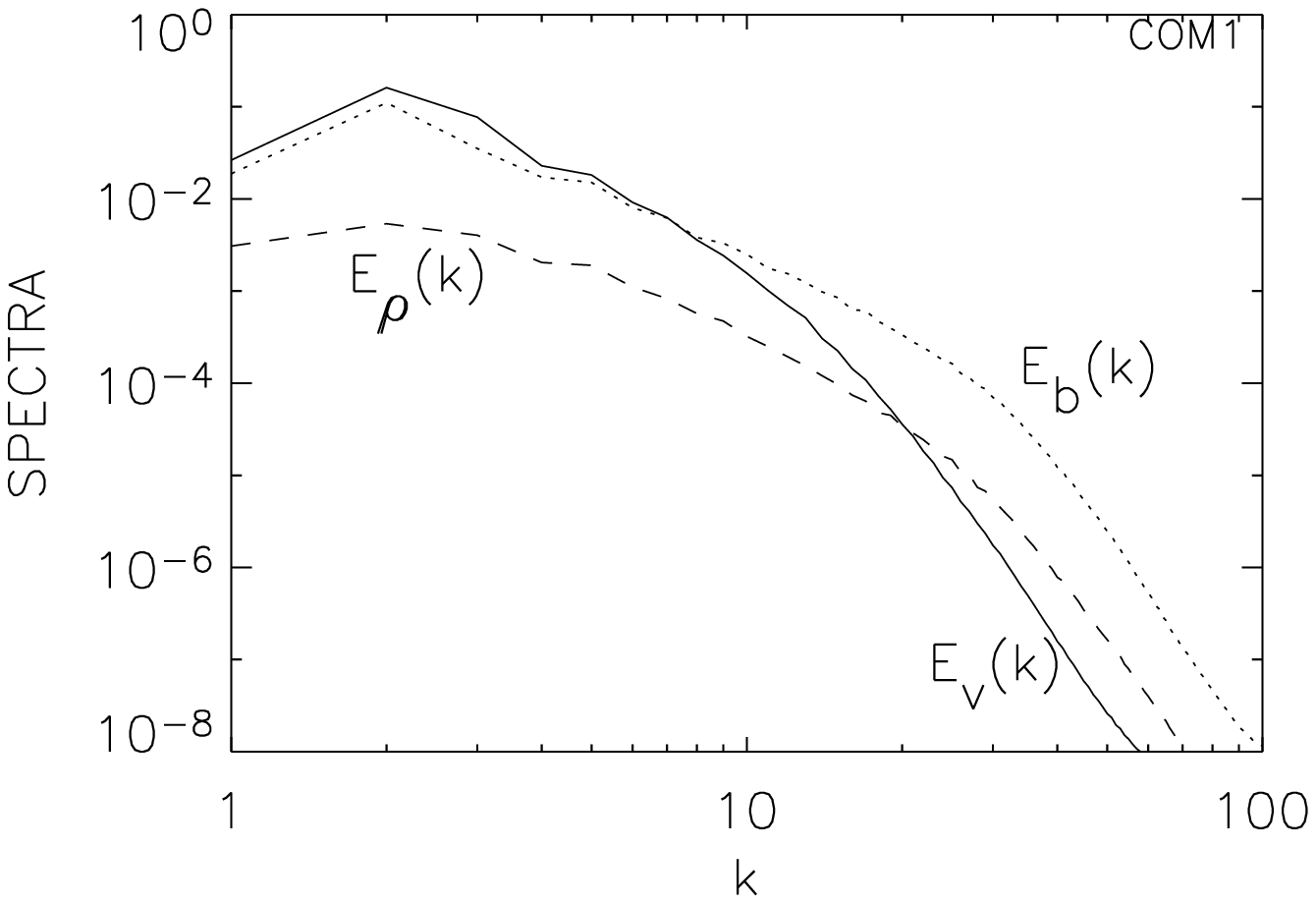}
  \caption{
    Spectra of runs without hyper-diffusion.
      We use the same physical viscosity ($\nu=0.015$) for velocity.
    (a: {\it left}) {}From an incompressible run with a physical
         magnetic diffusion (256PP-B$_0$1). 
         We use a small physical magnetic diffusion coefficient
        ($\eta=0.001$). 
        The kinetic spectrum declines quickly due to the large viscosity.
        The magnetic spectrum shows structures at small scales (i.e. $k>10$).
    (b: {\it right}) From a compressible MHD run (COM1). 
        We use numerical diffusion.
        The rms Mach number is $\sim$0.5.
        Due to the large viscosity ($\nu=0.015$), 
        the kinetic spectrum drops quickly for $k\gtrsim 7$.
        The magnetic spectrum shows structures at small scales (i.e. $k>10$).
        Note that the shape of the density spectrum ($E_{\rho}$) 
        is similar 
        to the magnetic spectrum. 
        Figure (b) is from Cho \& Lazarian (2003a,b).
}
\label{fig_sp_pp_com}
\end{figure*}

\begin{figure*}
  \includegraphics[width=0.32\textwidth]{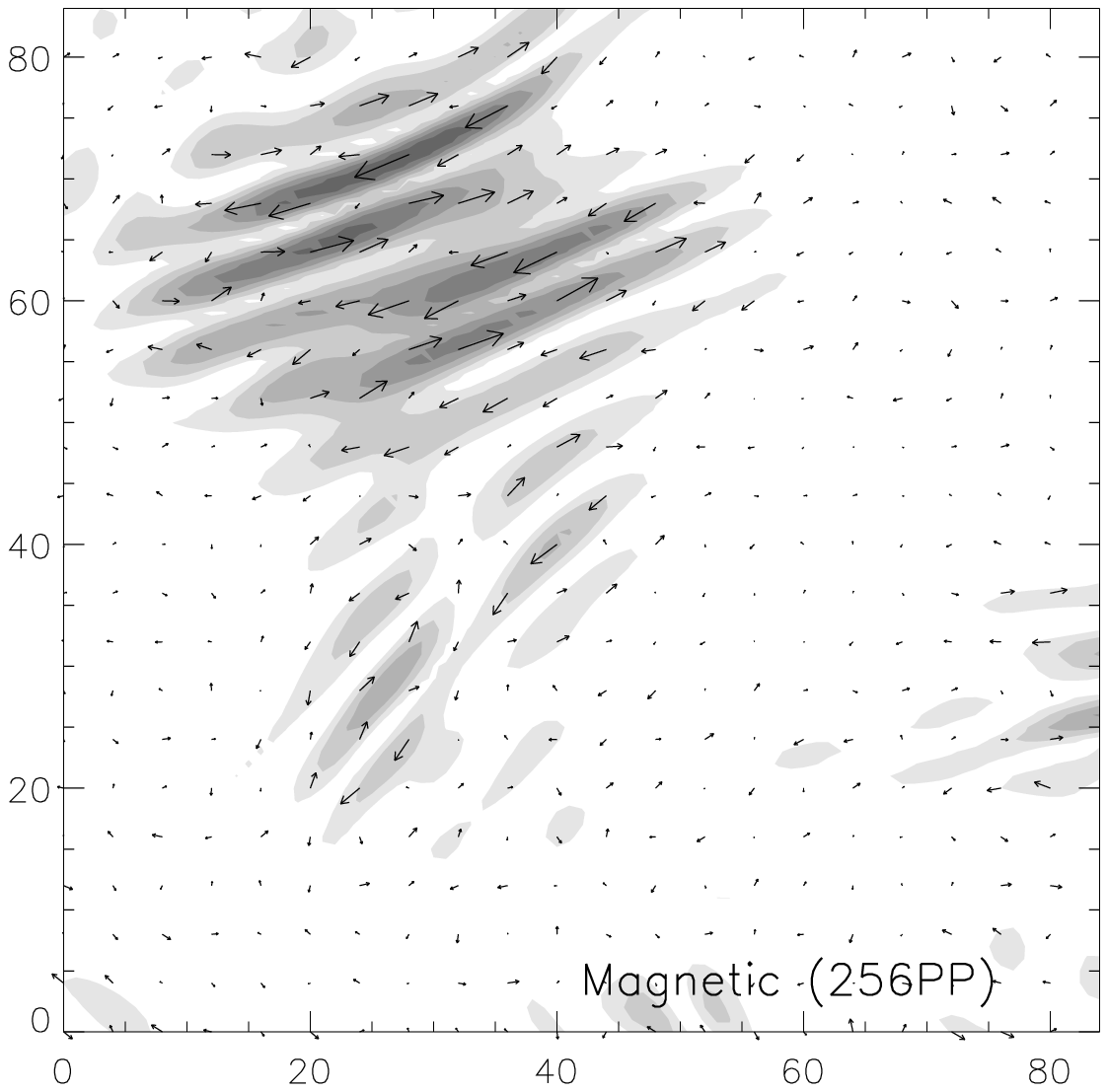}
\hfill
  \includegraphics[width=0.32\textwidth]{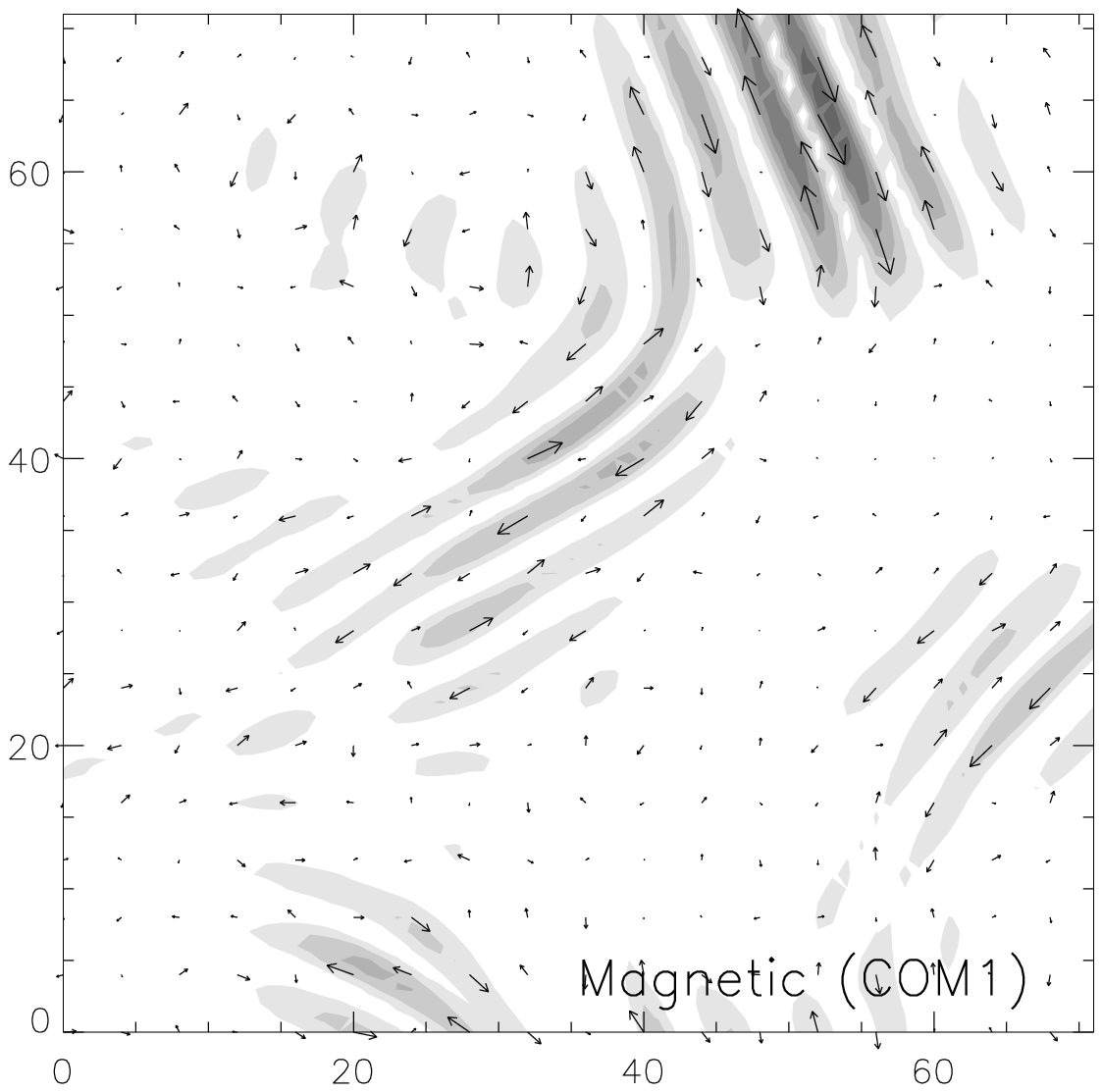}
\hfill
  \includegraphics[width=0.32\textwidth]{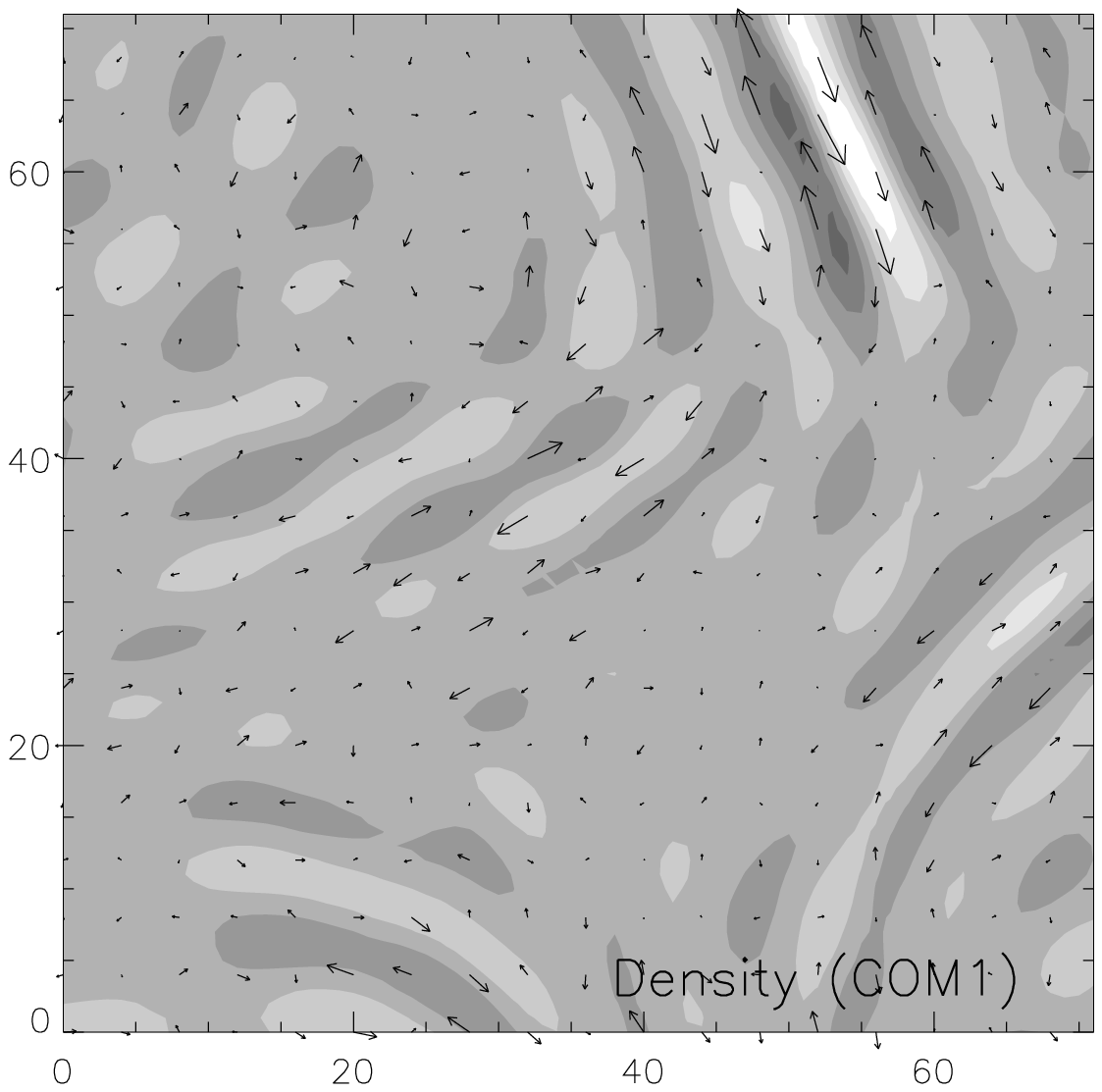}
  \caption{
      Intermittent small scale structures in a plane perpendicular
      to the mean magnetic field.
    (a: {\it left}) From an incompressible run with 
       a physical magnetic diffusion (256PP-B$_0$1).
        Small scale ($k>20$) magnetic field structures show intermittency.
    (b: {\it middle}) From a compressible run (COM1).
        Small scale ($k>20$) magnetic field structures.
    (c: {\it right}) From COM1. Small scale ($k>20$) density structures. 
}
\label{fig_conto_pp_com}
\end{figure*}

In Figure \ref{fig_sp_pp_com}(a), we see that, 
compared with its kinetic counterpart, the magnetic spectrum has
%%%-------------------
%structures at 
more power at
%%%-------------------
$k > 10$. Although we cannot say much about the spectrum of
the magnetic fluctuations at scales smaller than the
viscous cutoff ($k_c \sim 7$), the existence of the new regime is evident in
Figure \ref{fig_sp_pp_com}(a). 
A visual inspection of the magnetic structures in  
Figure \ref{fig_conto_pp_com}(a) exhibit intermittent
structures similar to those seen in the case of hyper-diffusion.
We see that the bottle-neck effect is not responsible for 
intermittent magnetic structures at small scales.

%%%%%%%%%%%%%%%%%%%%%%%%%%%%%%%%%%%%%%%%%%%%%Fig begin
%%%%%%%%%%%%%%%%%%%%%%%%%%%%%%%%%%%%%%%%%%%%%end Fig

Further testing of the viscous damped regime was performed in
Cho \& Lazarian (2003a,b). There, using the compressible code
mentioned earlier, we calculated not only velocity and magnetic field strength, 
but also density fluctuations that arise from MHD turbulence in
the viscosity-damped regime. 
We present our results in Figure \ref{fig_sp_pp_com}(b),
side by side with a plot of the incompressible simulation results.
The similarity between the two are vivid. It is also evident
that the shallow spectrum of magnetic fluctuations results
in a shallow spectrum of density fluctuations. Theoretically
we expect the density spectrum to follow a $k^{-1}$ law.
Unfortunately, our low resolution does not allow us to test this
prediction at the moment.

Figure \ref{fig_conto_pp_com}(c) shows that the density also has intermittent structures
below the viscous cutoff.
The density fluctuations may be (anti-)correlated with magnetic structures
(see Figure \ref{fig_conto_pp_com}(b)).
However, we will postpone further discussion of this point 
until higher resolution runs become available.

%%%%%%%%%%%%%%%%%%%%%%%%%%%%%%%%%%%%%%%%%%%%%Fig begin
%%%%%%%%%%%%%%%%%%%%%%%%%%%%%%%%%%%%%%%%%%%%%end Fig

\subsection{Comparison with Observations}

In this paper, we have considered properties of ordinary and 
viscosity-damped MHD turbulence. It is important to test 
our predictions with observations.

{\it Indirect Evidence} ---
Existence of density structures at scales smaller than a fraction of
a parsec is particularly important for the small scale density structures
in the ISM.
We speculate that the power-law density tail might have some relation to 
the tiny-scale atomic structures (TSAS).
Heiles (1997) introduced the term TSAS
for the mysterious
H~I absorbing structures on the scale from thousands to tens of
AU, discovered by Deiter, Welch \& Romney (1976). Analogs are observed
in NaI and CaII (Meyer \& Blades 1996; Faison \& Goss 2001; 
Andrews, Meyer \& Lauroesch 2001) and in molecular gas 
(Marscher, Moore \& Bania 1993). 
Recently Deshpande, Dwarakanath 
\& Goss (2000) 
analyzed channel maps of opacity fluctuations toward Cas A and Cygnus A.
They found 
that the amplitudes of density fluctuations at scales less than 0.1 pc 
are far larger than expected from extrapolation from larger scales, 
consistent with the existence of TSAS.

{\it Measuring velocity and density spectra} ---
Velocity channel analysis (VCA; Lazarian \& Pogosyan 2000) 
is a recent technique that
can extract velocity and density spectra.
The technique predicts that the power spectra of spectral line intensity vary 
when the thickness of velocity channels changes\footnote{
   Predictions by
   Lazarian \& Pogosyan (2000) have been confirmed
   through observations (Stanimirovic \& Lazarian 2001) and
   numerical calculations (Lazarian et al. 2001; Esquivel et al. 2003).}.
The compressible simulation in the previous subsection suggests that density 
may have a $k^{-1}$ spectrum below the viscous cutoff.
At these scales, velocity effects should be suppressed.
Deshpande et al. (2000) used absorption measurements 
to test the HI structure on subparsec scales and did
not see the variations of the channel map spectra as they
added their channels together. This meant that the velocity
effects were negligible at the scales they studied.
According to the VCA this meant that the shallow spectrum
of density $k^{-\alpha}$ ($\alpha < 1$) that they measured
corresponds to real density fluctuations. This spectrum may
arise from the new regime of MHD turbulence. Further
tests, including those which use different techniques
are necessary.

Information on the velocity statistics can also be obtained using centroids
of velocity (see Munch 1958). Modified velocity centroids (MVCs) that
are not sensitive to density fluctuations and can therefore be
used for studies of velocity statistics were proposed in Lazarian
\& Esquivel (2003).  They argued
that combining the VCA and MVCs it is possible to get reliable measures of the
underlying velocity spectra. It will be important to analyze the Deshpande
et al. data using MVCs.

{\it Studies of anisotropy} ---
Testing scale-dependent anisotropy is a challenging problem, because
scale-dependent anisotropy is averaged away
when we observe turbulence from outside.
The scale-dependent anisotropy is revealed only in a
local frame of reference whose parallel axis is parallel to
the local mean magnetic field (Cho \& Vishniac 2000; CLV02a).
Nevertheless, we can still study global anisotropy with 
3D maps using velocity centroids (Lazarian, Pogosyan \&
Esquivel 2002) or 2D maps obtained
from the Position-Position-Velocity data (Esquivel et al. 2003).

{\it Higher order statistics} ---
Scale-dependent intermittency is hard to study using observational
data. Usually the noise increases rapidly as we go to higher order statistics.
However, high order density
structure function scalings have been used to study
interstellar turbulence by Padoan et al. (2003b) and 
Padoan, Cambresy, \& Langer (2003). Further studies of the
problem are obviously required.

\section{Summary}
We have compared ordinary MHD turbulence and viscosity-damped MHD turbulence.
They are different in many ways - in their spectra, in the degree of anisotropy and
its scaling with length, and in the nature of intermittency in the turbulent cascade.
\begin{enumerate}
\item The spectra of ordinary strong MHD turbulence are compatible with 
Kolmogorov's $k^{-5/3}$ spectrum.
The magnetic spectra in viscosity-damped turbulence show a flatter spectrum:
$E(k) \sim k^{-1}$.

\item Eddies in ordinary MHD turbulence show scale-dependent anisotropy
consistent with the Goldreich \& Sridhar model.
Eddies in viscosity-damped MHD turbulence show extremely anisotropic
structures - their parallel size does not decrease significantly with the perpendicular
eddy size.

\item Intermittency of magnetic field strength
in ordinary turbulence is not scale-dependent.
Intermittency in viscosity-damped turbulence is scale-dependent:
smaller scales are more intermittent.

\item
The velocity and magnetic fields show different scaling exponents for
high order structure functions.

\item
Scaling exponents for the magnetic field in viscosity-damped MHD turbulence
show very little change as order of the structure changes.  They are generally
small and change from positive at $p=1$ to negative values at moderate $p$.

\end{enumerate}

\acknowledgements
J.C. thanks Peter Goldreich for clarifying the idea of calculating $k_{\|}$.
A.L.  acknowledges the support of NSF Grant AST-0125544.
E.V. acknowledges the support of NSF Grant AST-0098615.
This work was partially supported by National Computational Science
Alliance under AST000010N and  AST010011N and
utilized the NCSA SGI/CRAY Origin2000.

%%%%%%%%%%%%%%%%%%%%%%%%%%%%%%%%%%%%%%%%%%%%%%%%%%%%%%%%%%%%%%%%%%%%%
%\clearpage
%%%%%%%%%%%%%%%%%%%%%%%%%%%%%%%%%%%%%%%%%%%%%%%%%%%%%%%%%%%%%%%%%%%%%
\begin{deluxetable}{cccclr}
%\footnotesize
\tablecaption{Simulations}
\tablewidth{0pt}
\tablehead{
\colhead{Run \tablenotemark{a}} & \colhead{$N^3$} & \colhead{$\nu$} & \colhead{$\eta$} &
\colhead{$B_0$} & \colhead{Note}
}
\startdata

384H8H8-$B_0$1 & $384^3$ & hyper& hyper ($\nabla^{16}$)& 1& incompressible \\
384PH3-$B_0$1  & $384^3$ & .015 & hyper ($\nabla^{6}$) & 1& incompressible \\
256PH8-$B_0$0.5 & $256^3$& .06  & hyper ($\nabla^{16}$)& .5& incompressible \\
256PP-$B_0$1   & $256^3$ & .015 & .001                 & 1& incompressible  \\
COM1           & $216^3$ & .015 & numerical            & 1& compressible  \\
COM2           & $256^3$ & numerical & numerical       & 0& compressible  \\

\enddata

\tablenotetext{a}{ We use the notation 384XY-$B_0$Z (or 256XY-$B_0$Z),
where 384 (or 256) refers to the number of grid points in each spatial
direction; X, Y = P, H3, H8 refers to physical or hyper-diffusion 
(and its power);
Z=0.5, 1 refers to the strength of the external magnetic field.
    }
%\tablenotetext{b}{ Same as the run 256H-B$_0$1 in Cho \& Vishniac (2000b).}

\end{deluxetable}
%%%%%%%%%%%%%%%%%%%%%%%%%%%%%%%%%%%%%%%%%%%%%%%%%%%%%%%%%%%%%%%%%%%%%

\end{document}